\documentclass[reqno]{article}
\usepackage{amssymb,amsmath,cite,delarray,epsfig,hhline}
\newcommand{\lsim} {\buildrel < \over {_\sim}}
\newcommand{\gsim} {\buildrel > \over {_\sim}}

\begin{document}

\title{\textbf{A Unified Picture with Neutrino As a Central Feature }\thanks{Invited talk presented at the XI International
Workshop on ``Neutrino Telescopes'' held at Venice, February
21--25, 2005, to appear in the proceedings.}}
\author{Jogesh C. Pati\\
Department of Physics, University of Maryland, College Park, MD
20740 USA} \maketitle \setcounter{footnote}{0}

\vskip.25in
\begin{abstract}
In the first part of this talk it is discussed why observed
neutrino oscillations (which suggest the existence of right-handed
neutrinos with certain Dirac and Majorana masses) seem to select
out the route to higher unification based on the symmetry
SU(4)-color. This in turn selects out the effective symmetry in 4D
near the GUT/string scale to be either SO(10) or minimally
$G(224)= SU(2)_L\times SU(2)_R \times SU(4)^c$. The same
conclusion is reached by the likely need for leptogenesis as the
means for baryogenesis and also by the success of certain fermion
mass-relations including $m_b(M_{GUT})\approx m_\tau$, together
with $m(\nu^\tau)_{Dirac}\approx m_{top}(M_{GUT})$.

In the second part, an attempt is made to provide a unified
picture of a set of diverse phenomena based on an effective G(224)
symmetry or SO(10), possessing supersymmetry. The phenomena in
question include: (a) fermion masses and mixings, (b) neutrino
oscillations, (c) CP non-conservation, (d) flavor violations in
quark and lepton sectors, as well as (e) baryogenesis via
leptogenesis. Including SM and SUSY contributions, the latter
being sub-dominant, the framework correctly accounts for $\Delta
m_K,~ \Delta m_{B_d}, ~ S(B_d\to J/\psi K_s)$ {\it and }
$\epsilon_K$, and predicts $S(B_d\to \phi K_s)$ to be in the range
+(0.65--0.73), close to the SM-prediction. It also quite plausibly
accounts for the observed baryon excess $Y_B\approx 10^{-10}$.
Furthermore the model predicts enhanced rates for $\mu\to e\gamma,
~\tau\to \mu\gamma$ and $\mu N\to e N$ and also measurable
electric dipole moment for the neutron. Expectations arising
within the same framework for proton decay are summarized at the
end. It is stressed that the two notable missing pieces of the
framework are supersymmetry and proton decay. While search for
supersymmetry at the LHC is eagerly awaited, that for proton decay
will need the building of a megaton-size detector.
\end{abstract}

\newpage
{\large

\section{Introduction}
Since the discoveries (confirmations) of the atmospheric \cite{sk}
and solar neutrino oscillations \cite{sno,Bahcall}, the neutrinos
have emerged as being among the most effective probes into the
nature of higher unification. Although almost the feeblest of all
the entities of nature, simply by virtue of their tiny masses,
they seem to possess a subtle clue to some of the deepest laws of
nature pertaining to the unification-scale and (even more
important) to the nature of the unification-symmetry. In this
sense the neutrinos provide us with a rare window to view physics
at truly short distances. As we will see, these turn out to be as
short as about $10^{-30}$ cm. In addition,it appears most likely
that the origin of their tiny masses may be at the root of the
origin of matter-antimatter asymmetry in the early universe. In
short, the neutrinos may be crucial to shedding light not only on
unification but also on our own origin!

The main purpose of this talk  would be two-fold.  First I discuss
in the next section the issue of the choice of the effective
symmetry in 4D. Here, I explain why (a) observed neutrino
oscillations, (b) the likely need for leptogenesis as the means
for baryogenesis \cite{Yanagida,PatiLepto}, and (c) the success of
certain fermion mass relations, {\it together}, seem to select out
the route to higher unification based on the symmetry SU(4)-color
\cite{JCPAS,PS}. The effective symmetry near the GUT/string scale
in 4D should thus be either SO(10) \cite{SO(10)}, or minimally
$G(224)= SU(2)_L\times SU(2)_R \times SU(4)^c$ \cite{PS}, as
opposed to other alternatives. The second part of my talk is based
on recent works on fermion masses and neutrino oscillations
\cite{BPW}, and CP and flavor violations \cite{BPR,LFV}, all
treated within a promising SO(10)/G(224) framework. The purpose of
this second part is to present a {\it unified description} of a
set of diverse phenomena, including:
\begin{itemize}
\item Fermion masses and mixings \item Neutrino oscillations \item
CP non-conservation \item Flavor violations (in quark {\it and}
lepton sectors), \item Baryogenesis via leptogenesis, and \item
Proton Decay.
\end{itemize}

As it turns out, the neutrino plays a central role in arriving at
this unified picture. My goal here will be to exhibit that the
first five phenomena hang together neatly, in accord with
observations, within {\it a single predictive framework}, based on
an effective symmetry in 4D which is either SO(10) or G(224).

As we will see, the predictions of the framework not only account
for many of the features of the five phenomena listed above
(including the smallness of $V_{cb}$, the near maximality of
$\Theta^\nu_{23}$, $m_b(m_b)$, $\Delta m^2(\nu_2-\nu_3)$,
$\epsilon_K$, $S(B_d \to J/\psi K_S)$, baryon asymmetry $Y_B$, and
more), but also include features involving CP and flavor
violations (such as edm, the asymmetry parameter $S(B_d \to \phi
K_S)$ and $\mu\to e\gamma$) which can clearly test the framework
on many fronts.

To set the background for this discussion I first remark in the
next section on the choice of the effective symmetry in 4D and the
need for SU(4)-color. In this connection, I also clarify the
historical origin of some of the concepts that are common to both
G(224) and SO(10) and are now crucial to an understanding of
neutrino masses and implementing baryogenesis via leptogenesis. In
the following section, I briefly review the
SO(10)/G(224)-framework proposed in Ref. \cite{BPW} for
considerations of fermion masses and neutrino oscillations, and in
the subsequent sections discuss the issues of CP and flavor
violations \cite{BPR,LFV} as well as baryogenesis via leptogenesis
\cite{PatiLepto}, within the same framework. Expectations for
proton decay are noted at the end.

\section{On the choice of the Effective Symmetry in 4D: The need for
SU(4)-color}

The idea of grand unification was motivated \cite{JCPAS,PS,SU(5)}
by the desire to explain (a) the observed quantum numbers of the
members of a family, and (b) quantization of electric charge on
the one hand, and simultaneously to achieve (c) unification of
quarks and leptons and (d) a unity of the basic forces on the
other hand. While these four, together with the observed gauge
coupling unification \cite{GQW} , still provide the strongest
support -- on aesthetic and empirical grounds-- in favor of grand
unification, they leave open the question of the choice of the
effective symmetry G in 4D near the GUT scale which achieves these
four goals.

For instance, should the symmetry group G be of rank 4, that is
SU(5) \cite{SU(5)}, which is devoid of SU(4)-color? Or, should G
possess SU(4)-color and thus minimally be SO(10) of rank 5, or
even $E_6$ \cite{E6} of rank 6? Or, should G be a string-derived
semi-simple group G(224) $\subset$ SO(10), still of rank 5? Or,
should G be [SU(3)]$^3\subset E_6$, of rank 6, but devoid of
SU(4)-color?

An answer to these questions that helps select out the effective
symmetry G in 4D is provided, however, if together with the four
features (a)--(d) listed above, one folds in the following
three:\\
\fbox{\(
\begin{array}{l}
\mbox{(e) Neutrino oscillations} \\ \mbox{(f) The
likely need for leptogenesis as the means for baryogenesis, and} \\
\mbox{(g) The success of certain fermion mass relations noted
below}
\end{array}\)}

One can argue \cite{JCPKeK} that the last three features, together
with the first four listed above, clearly suggest that the
standard model symmetry very likely emerges, near the GUT-scale
$M_U \sim 2\times 10^{16}$ GeV, from the spontaneous breaking of a
higher gauge symmetry G that should possess the symmetry
SU(4)-color \cite{PS}. The relevant symmetry in 4D could then
maximally be SO(10) (possibly even $E_6$ \cite{E6}) or minimally
the symmetry G(224); either one of these symmetries may be viewed
to have emerged in 4D \cite{stringSO(10),stringG(224)} from a
string/M theory near the string scale $M_{st}\gsim M_{GUT}$
\footnote{The relative advantage of an effective string-derived
SO(10) over a G(224)-solution and vice versa have been discussed
in detail in \cite{JCPKeK}. Briefly speaking, for the case of a
string derived G(224)-solution, coupling unification being valid
near the string scale, one needs to assume that the string scale
is not far above the GUT scale ($M_{st}\approx (2-3) M_{GUT}$,
say) to explain observed gauge coupling unification. While such a
possibility can well arise in the string theory context
\cite{Witten96}, for an SO(10)-solution, coupling unification at
the GUT-scale is ensured regardless of the gap between string and
GUT-scales. The advantage of a G(224)-solution over an SO(10)
solution is, however, that doublet-triplet splitting (DTS) can
emerge naturally for the former in 4D through the process of
string compactification (see Ref. \cite{stringG(224)}), while for
an SO(10)-solution this feature is yet to be realized. As we will
see, SO(10) and G(224) share many common advantages, aesthetic and
practical, in particular as regards an understanding of fermion
masses, neutrino oscillations and baryogenesis via leptogenesis;
but they can be distinguished empirically through phenomena
involving CP and flavor violations as well as proton decay.}. The
theory thus described should of course possess weak scale
supersymmetry so as to avoid unnatural fine tuning in Higgs mass
{\it and} to ensure gauge coupling unification.

To see the need for having SU(4)-color as a component of the
higher gauge symmetry, it is useful to recall the family-multiplet
structure of G(224), which is retained by SO(10) as well. The
symmetry G(224), subject to left-right discrete symmetry which is
natural to G(224), organizes members of a family into a single
left-right self-conjugate multiplet
($\rm{F_L^e}\bigoplus\rm{F_R^e}$) given by \cite{PS}:
\begin{eqnarray}
\label{eq:fermion}
\begin{array}{c}
{\rm F_{L,R}^e}=\left[
\begin{array}{cccc}
{\rm u_r}&{\rm u_y}&{\rm u_b}&\mathbf{\nu_e}\\{\rm d_r}&{\rm
d_y}&{\rm d_b}&{\rm e^-}
\end{array}\right]_{\rm L,R}
\end{array}
\end{eqnarray}

The multiplets $\rm{F_L^e}$ and $\rm{F_R^e}$ are left-right
conjugates of each other transforming respectively as {\bf (2, 1,
4)} and {\bf (1, 2, 4)} of G(224); likewise for the muon and the
tau families. Note that {\it each family of G(224), subject to
left-right symmetry, must contain sixteen two-component objects}
as opposed to fifteen for SU(5) \cite{SU(5)} or the standard
model. While the symmetries $SU(2)_{L,R}\subset G(224)$ treat each
column of $\rm{F_{L,R}^e}$ as doublets, the symmetry SU(4)-color
unifies quarks and leptons by treating each row of $\rm{F_L^e}$
and $\rm{F_R^e}$ as a quartet. Thus SU(4)-color treats the left
and right-handed neutrinos ($\nu_L^e$ and $\nu_R^e$) as the {\it
fourth color-partners} of the left and right-handed up quarks
(u$_{\rm L}$ and u$_{\rm R}$) respectively. Here in lies the
distinctive feature of SU(4)-color. {\it It necessitates the
existence of the RH neutrino ($\nu_R^e$) on par with that of the
RH up quark (u$_{\rm R}$) by relating them through a gauge
symmetry transformation;} and likewise for the mu and the tau
families. As we will see, this in turn leads to some very
desirable fermion mass relations for the third family that help
distinguish it from alternative symmetries. {\it An accompanying
characteristic of SU(4)-color is that it also introduces $B-L$ as
a local symmetry} \cite{PS}. This in turn plays a crucial role in
protecting the Majorana masses of the right-handed neutrinos from
acquiring Planck-scale values.

In anticipation of sections. 3, 4 and 7 where some of the
statements made below will become clear, I may now state the
following. The need for SU(4)-color (mentioned above) arises
because it provides the following desirable features:\\
$\begin{array}{ll}
(1){\rm\ RH\ neutrino\ (\nu_R^i)\ as\ an\ essential} & {\rm ~~~~~~Needed\ to\ implement\ the\ seesaw\ mechanism}\\
{\rm ~~~~\ member\ of\ each\ family}&{\rm ~~~~~~and\
leptogenesis\ (see\ Secs.\ 3\ and\ 7).}\\

(2){\rm\ B-L\ as\ a\ local\ symmetry} & {\rm ~~~~~~Needed\ to\
protect\
\nu_R 's\ from\ acquiring\ Planck\ }\\
 &  {\rm ~~~~~~scale\ masses\ and\ to\ set\ M(\nu_R^i)\propto M_{B-L}\sim M_{GUT}.}\\
\end{array}$

\noindent \ (3) Two simple mass relations for

\noindent \ \ \ \ \ \ the\  3rd\ family:
\\
$\begin{array}{ll}
{\rm ~~~~ (a)\ m(\nu_{\rm Dirac}^\tau)}\ \approx\ {\rm m_{\rm top}(M_{\rm GUT})} &~~~~~~~~~~~{\rm Needed\ for\ success\ of\ seesaw\ ( see\ section\ 3).}\\
{\rm~~~~ (b)\ m_{\rm b}(M_{\rm GUT})}\ \approx\ {\rm m_\tau}
&~~~~~~~~~~{~\rm Empirically\ successful.}
\end{array}$
\\
\\
These three ingredients ((1), (2) and (3a)), together with the
SUSY unification-scale, are indeed crucial (see sections 3 and 4)
to an understanding of the neutrino masses via the seesaw
mechanism \cite{seesaw}. The first two ingredients are important
also for implementing baryogenesis via leptogenesis
\cite{Yanagida,PatiLepto} (see section 7). {\it Hence the need for
having SU(4)-color as a component of the unification symmetry
which provides all four ingredients.}

By contrast SU(5), devoid of SU(4)-color, does not provide the
ingredients of (1), (2) and (3a) (though it does provide (3b));
hence it does not have a natural setting for understanding
neutrino masses and implementing baryogenesis via leptogenesis
(see discussion in section 4 and especially footnote 2).
Symmetries like $G(2213)=SU(2)_L\times SU(2)_R\times
U(1)_{B-L}\times SU(3)^c$\cite{G(2213)} and $[SU(3)]^3$
\cite{[SU(3)]^3} provide (1) and (2) but neither (3a) nor (3b),
while flipped $SU(5)'\times U(1)'$ \cite{flipSU(5)} provides (1),
(2) and (3a) but not (3b). In summary, {\it the need for the
combination of the four ingredients (1), (2), (3a) and (3b) seems
to select out the route to higher unification based on
SU(4)-color}, and thereby as mentioned above an effective symmetry
like G(224) or SO(10) being operative in 4D near the string scale.

At this point, an intimate link between $SU(4)$-color and the
left-right symmetric gauge structure $SU(2)_L\times SU(2)_R$ is
worth noting. Assuming that $SU(4)^c$ is gauged and demanding an
explanation of quantization of electric charge lead one to gauge
minimally the left-right symmetric flavor symmetry $SU(2)_L\times
SU(2)_R$ (rather than $SU(2)_L\times U(1)_{I_{3R}}$). The
resulting minimal gauge symmetry that contains $SU(4)$-color and
explains quantization of electric charge is then
$G(224)=SU(2)_L\times SU(2)_R\times SU(4)^c$ \cite{PS}. With
SU(4)-color being vectorial, such a symmetry structure (as also
G(2213) which is a subgroup of G(224)) in turn naturally suggests
the attractive idea that L--R discrete symmetry and thus parity
(i.e. F$_{\rm L}\leftrightarrow$ F$_{\rm R}$, W$_{\rm
L}\leftrightarrow$ W$_{\rm R}$ with g$_{\rm L}^{(0)}=$g$_{\rm
R}^{(0)}$) is preserved at a basic level and is broken only
spontaneously \cite{parity}. In other words, observed parity
violation is only a low-energy phenomenon which should disappear
at sufficiently high energies. {\it We thus see that the concepts
of SU(4)-color and left-right symmetry are intimately
inter-twined, through the requirement of quantization of electric
charge}.
\\
\\
\noindent{\bf A Historical Note: Advantages of G(224)}
\\
As a historical note, it is worth noting that the symmetry
$SU(4)$-color, and thereby the three desirable features listed
above, were introduced into the literature, as a step towards
higher unification, through the minimal symmetry G(224) \cite{PS},
rather than through SO(10) \cite{SO(10)}. The symmetry G(224)
(supplemented by L--R discrete symmetry which is natural to
G(224)) in turn brought a host of desirable features. Including
those mentioned above they are:

\noindent(i) Unification of all sixteen members of a family within
{\it one} left-right self-conjugate multiplet, with a neat
explanation of
their quantum numbers;\\ (ii) Quantization of electric charge;\\
(iii) Quark-lepton unification through $SU(4)$-color;\\ (iv)
Conservation of parity at a fundamental level\cite{parity};\\ (v) RH neutrino as a compelling member of each family;\\
(vi) $B-L$ as a local symmetry; and \\ (vii) The rationale for the
now successful mass-relations (3a) and (3b).

These seven features constitute the {\it hallmark} of G(224). Any
simple or semi-simple group that contains G(224) as a subgroup
would of course naturally possess these features. So does
therefore SO(10) which is the smallest simple group containing
G(224). Thus, as alluded to above, {\it all the attractive
features of SO(10), which distinguish it from SU(5) and are now
needed to understand neutrino masses and baryogenesis via
leptogenesis, were in fact introduced through the symmetry G(224)
\cite{PS}, long before the SO(10) papers appeared \cite{SO(10)}}.
These in particular include the features (i) as well as
(iii)--(vii). SO(10) of course preserved these features for
reasons stated above; it even preserved the family multiplet
structure of G(224) without needing additional fermions (unlike
$E_6$) in that the L--R conjugate {\bf 16-plet} ( = $F_L\bigoplus
F_R$) of G(224) precisely corresponds to the spinorial {\bf 16} (=
$F_L\bigoplus (F_R)^c$) of SO(10). Furthermore, with SU(4)-color
being vectorial, G(224) is anomaly-free; so also is SO(10).

SO(10) brought of course one added and desirable feature relative
to G(224)-- that is manifest coupling unification. Again, as a
historical note, it is worth mentioning that the idea of coupling
unification was initiated in \cite{JCPAS} and was first manifested
explicitly within a minimal model through the suggestion of SU(5)
in \cite{SU(5)}.

As mentioned before, believing in string unification, either
G(224) or SO(10) may be viewed to have its origin in a still
higher gauge symmetry (like $E_8$) in 10D. To realize the
existence of the right-handed neutrinos, $B-L$ as a local symmetry
and the fermion mass-relations (3a), which are needed for
understanding neutrino masses and implementing baryogenesis via
leptogenesis, I have argued that one needs $SU(4)^c$ as a
component of the effective symmetry in 4D, and therefore minimally
G(224) (or even G(214)) or maximally perhaps SO(10) in 4D near the
string scale. The relative advantages of G(224) over SO(10) and
vice versa as 4D symmetries in addressing the issues of
doublet-triplet splitting on the one hand and gauge coupling
unification on the other hand have been discussed in Ref.
\cite{JCPKeK} and briefly noted in footnote 1.

In the following sections I discuss how either one of these
symmetries G(224) or SO(10) link together fermion masses, neutrino
oscillations, CP and flavor violations and leptogenesis. As we
will see, while G(224) and SO(10) lead to essentially identical
results for fermion masses and neutrino oscillations, which are
discussed in the next two sections, they can be distinguished by
processes involving CP and/or flavor violations, which are
discussed in sections 5 and 6, and proton decay, discussed in
section 8.

\section{Seesaw and SUSY Unification with SU(4)-color}
The idea of the seesaw mechanism \cite{seesaw} is simply this.  In
a theory with RH neutrinos as an essential member of each family,
and with spontaneous breaking of $B-L$ and $I_{3R}$ at a high
scale ($M_{B-L}$), both already inherent in \cite{PS}, the RH
neutrinos can and generically will acquire a superheavy Majorana
mass ($M(\nu_R) \sim M_{B-L}$) that violates lepton number and
$B-L$ by two units. Combining this with the Dirac mass of the
neutrino ($m(\nu_{\mathrm{Dirac}})$), which arises through
electroweak symmetry breaking, one would then obtain a mass for
the LH neutrino given by
\begin{equation}
m(\nu_L) \approx m(\nu_{\mathrm{Dirac}})^2/M(\nu_R) \label{eq:4}
\end{equation}
which would be naturally super-light because $M(\nu_R)$ is
naturally superheavy.  This then provided a \emph{simple but
compelling reason} for the lightness of the known neutrinos.  In
turn it took away the major burden that faced the ideas of
$SU(4)$-color and left-right symmetry from the beginning. In this
sense, the seesaw mechanism was indeed the missing piece that was
needed to be found for consistency of the ideas of $SU(4)$-color
and left-right symmetry.

In turn, of course, the seesaw mechanism needs the ideas of
$SU(4)$-color and SUSY grand unification so that it may be
quantitatively useful.  Because the former provides (a) the RH
neutrino as a compelling feature (crucial to seesaw), and (b) the
Dirac mass for the tau neutrino accurately in terms of the top
quark mass (cf.\ feature (3a)), while the latter provides the
superheavy Majorana mass of the $\nu_{R}^{\tau}$ in terms of the
SUSY unification scale (see Sec.~4). Both these masses enter
crucially into the seesaw formula and end up giving the
\emph{right mass-scale} for the atmospheric neutrino oscillation
as observed. To be specific, $SU(4)$-color yields:
$m(\nu_{\mathrm{Dirac}}^{\tau}) \approx m_{\mathrm{top}}(M_U)
\approx 120 \mathrm{\ GeV}$; and the SUSY unification scale,
together with the protection provided by $B-L$ that forbids
Planck-scale contributions to the Majorana mass of
$\nu_{R}^{\tau}$, naturally yields: $M(\nu_{R}^{\tau}) \sim
M_{GUT}^2/M \sim 4\times 10^{14}\mathrm{\ GeV}(1/2\mbox{--}2)$,
where $M\sim 10^{18}$ GeV $(1/2\mbox{--}2)$ [cf. Sec.~4]. The
seesaw formula (without 2-3 family mixing) then yields:
\begin{eqnarray}
m(\nu_{L}^{3}) \approx (120\mathrm{\ GeV})^{2}/(4\times
10^{14}\mathrm{\ GeV}(1/2\mbox{--}2) )\approx (1/28\mathrm{\
eV})(1/2\mbox{--}2))
\end{eqnarray}

\noindent With hierarchical pattern for fermion mass-matrices (see
Sec. 4), one necessarily obtains $m(\nu_{L}^{2})\ll
m(\nu_{L}^{3})$ (see section 4), and thus $\sqrt{\Delta
m_{23}^2}\approx m(\nu_{L}^{3})\sim 1/28 \mathrm{\
eV}(1/2\mbox{--}2)$. This is just the right magnitude to go with
the mass scale observed at SuperK \cite{sk}!

\emph{Without an underlying reason as above for at least the
approximate values of these two vastly differing mass-scales ---
$m(\nu_{\mathrm{Dirac}}^{\tau})$ and $M(\nu_{R}^{\tau})$ --- the
seesaw mechanism by itself would have no clue, quantitatively, to
the mass of the LH neutrino.}  In fact it would yield a rather
arbitrary value for $m(\nu_{L}^{\tau})$, which could vary quite
easily by more than 10 orders of magnitude either way around the
observed mass scale. This would in fact be true if one introduces
the RH neutrinos as a singlet of the SM or of SU(5).\footnote{To
see this, consider for simplicity just the third family.  Without
$SU(4)$-color, even if a RH two-component fermion $N$ (the
analogue of $\nu_R$) is introduced by hand as a \emph{singlet} of
the gauge symmetry of the SM or $SU(5)$, \emph{such an $N$ by no
means should be regarded as a member of the third family, because
it is not linked by a gauge transformation to the other fermions
in the third family}.  Thus its Dirac mass term given by
\(m(\nu_{\mathrm{Dirac}}^{\tau})[ \bar{\nu}_{L}^{\tau} N + h.c.
]\) is completely arbitrary, except for being bounded from above
by the electroweak scale $\sim 200 \mbox{\ GeV}$. In fact a priori
(within the SM or $SU(5)$) it can well vary from say 1~GeV (or
even 1~MeV) to 100~GeV. Using Eq.~(\ref{eq:4}), this would give a
variation in $m_(\nu_L)$ by at least four orders of magnitude if
the Majorana mass $M(N)$ of $N$ is held fixed.  Furthermore, $N$
being a singlet of the SM as well as of $SU(5)$, the Majorana mass
$M(N)$, unprotected by $B-L$, could well be as high as the Planck
or the string scale (\(10^{18}\mbox{--}10^{17}\mbox{\ GeV}\)), and
as low as say 1~TeV; this would introduce a further arbitrariness
by fourteen orders of magnitude in $m(\nu_L)$. Such arbitrariness
both in the Dirac and in the Majorana masses, is drastically
reduced, however, once $\nu_R$ is related to the other fermions in
the family by an $SU(4)$-color gauge transformation and a SUSY
unification is assumed.}

In short, the seesaw mechanism needs the ideas of SUSY unification
and $SU(4)$-color, and of course vice-versa; \emph{together} they
provide an understanding of neutrino masses as observed.
Schematically, one thus finds:
\begin{equation}
\begin{array}{rcl}
\fbox{\(
\begin{array}{c}
\mbox{SUSY UNIFICATION} \\ \mbox{WITH $SU(4)$-COLOR}
\end{array}\)} & \oplus & \fbox{SEESAW} \\
 & \Downarrow & \\
m(\nu_{L}^{3}) & \sim & 1/10 \mbox{\ eV}.
\end{array}
\label{eq:scheme}
\end{equation}

In summary, as noted in section 2, the agreement of the expected
$\sqrt{\Delta m_{23}^2}$ with the observed SuperK value clearly
seems to favor the idea of the seesaw and select out the route to
higher unification based on supersymmetry and SU(4)-color, as
opposed to other alternatives.

I will return to a more quantitative discussion of the mass scale
and the angle associated with the atmospheric neutrino
oscillations in Sec.~4.

\section{Fermion Masses and Neutrino Oscillations in
G(224)/SO(10): A Review of the BPW framework} Following
Ref.~\cite{BPW}, I now present a simple and predictive pattern for
fermion mass-matrices based on $SO(10)$ or the
$G(224)$-symmetry.\footnote{I will present the Higgs system for
$SO(10)$.  The discussion would remain essentially unaltered if
one uses the corresponding $G(224)$-submultiplets instead.} One
can obtain such a mass mass-matrix for the fermions by utilizing
only the minimal Higgs system that is used also to break the gauge
symmetry $SO(10)$ to \(SU(3)^{c} \times U(1)_{em}\).  It consists
of the set:
\begin{equation}
H_{\mathrm{minimal}} = \left\{ \mathbf{45_{H}}, \mathbf{16_{H}},
\overline{\mathbf{16}}_{\mathbf{H}}, \mathbf{10_{H}} \right\}
\label{Hmin}
\end{equation}
Of these, the VEV of \(\left\langle \mathbf{45_{H}} \right\rangle
\sim M_{U}\) breaks $SO(10)$ in the B-L direction to \(G(2213) =
SU(2)_{L} \times SU(2)_{R} \times U(1)_{B-L} \times SU(3)^{c}\),
and those of \(\left\langle \mathbf{16_{H}} \right\rangle =
\left\langle \overline{\mathbf{16}}_{\mathbf{H}} \right\rangle\sim
M_U\) along \(\left\langle \tilde{\bar{\nu}}_{RH} \right\rangle\)
and \(\left\langle \tilde{\nu}_{RH} \right\rangle\)\ break
$G(2213)$ into the SM symmetry $G(213)$ at the unification-scale
$M_{U}$. Now $G(213)$ breaks at the electroweak scale by the VEVs
of \(\left\langle \mathbf{10_{H}} \right\rangle\)  and of the EW
doublet in \(\left\langle \mathbf{16_{H}} \right\rangle\) to
\(SU(3)^{c} \times U(1)_{em}\).

Before discussing fermion masses and mixings, I should comment
briefly on the use of the minimal Higgs system noted  above as
opposed to large-dimensional tensorial multiplets of SO(10)
including ($126_H,~\overline{126_H}$), $210_H$ and possibly
$120_H$, which have been used widely in the literature
\cite{SO(10)126} to break SO(10) to the SM symmetry and give
masses to the fermions. We have preferred to use the
low-dimensional Higgs multiplets ($45_H,~ 16_H$ and
$\overline{16_H}$) rather than the large dimensional ones like
($126_H,~\overline{126_H}$, $210_H$ and possibly $120_H$) in part
because these latter tend to give too large GUT-scale threshold
corrections to $\alpha_3(m_Z)$ from split sub-multiplets
(typically exceeding 15--20\% with either sign), which would
render observed gauge coupling unification fortuitous (see
Appendix D of Ref. \cite{BPW} for a discussion on this point). By
contrast, with the low-dimensional multiplets ($45_H,~ 16_H$ and
$\overline{16_H}$), the threshold corrections to $\alpha_3(m_Z)$
are tamed and are found, for a large range of the relevant
parameters, to have the right sign and magnitude (nearly -5 to
-8\%) so as to account naturally for the observed gauge coupling
unification.

Another disadvantage of $126_H$, which contributes to EW symmetry
breaking through its ($2,~2,~15 $) component of G(224), is that it
gives $B-L$ dependent contribution to family-diagonal fermion
masses. Such a contribution, barring adjustment of parameters
against the contribution of \(\left\langle 10_{H} \right\rangle\),
could in general make the success of the relation $m_b(GUT)\approx
m_{\tau}$ fortuitous. By contrast, the latter relation emerges as
a robust prediction of the minimal Higgs system ($ 45_H,16_H,
\overline {16}_H \mbox{ and } 10_H $), subject to a hierarchical
pattern, because the only $(B-L)$ dependent contribution in this
case can come effectively through \(\left\langle 10_{H}
\right\rangle\)\(\left\langle 45_{H} \right\rangle\)/$M$ which is
family-antisymmetric and cannot contribute to diagonal entries
(see below) \cite{BPW}.

One other issue involves the question of achieving doublet-triplet
splitting by a natural mechanism as opposed to that of fine
tuning, and incorporating the associated GUT-scale threshold
correction to $\alpha_3(m_Z)$. For the case of ($45_H,~ 16_H$,
$\overline{16_H}$ and $10_H$), there exists a simple mechanism
which achieves the desired splitting naturally with the
introduction of an extra $10'_H$ \cite{DimWil}, and the effect of
this splitting on GUT-scale threshold correction to
$\alpha_3(m_Z)$ has been evaluated in \cite{BPW} to conform with
natural coupling unification on the one hand and the limit on
proton lifetime on the other hand. To the best of my knowledge, an
analogous study for the system involving
($126_H,~\overline{126_H}$) has not been carried out as yet.

Balancing against these advantages of the minimal Higgs system,
the large-dimensional system ($ 126_H, \overline {126}_H, 210_H
\mbox{ and possibly~} 120 _H $) has an advantage over the minimal
system, because $126$ and $\overline{126}$ break $B-L$ by two
units and thus automatically preserve the familiar R-parity =
$(-1)^{3(B-L)+2S}$. By contrast, $16$ and $\overline{16}$ break
$B-L$ by one unit and thereby break the familiar R-parity. This
difference is, however, not really significant, because for the
minimal system one can still define consistently a matter-parity
(i.e. $16_i\to -16_i, ~16_H \to 16_H, ~\overline{16_H} \to
\overline{16_H}, ~45_H \to 45_H, ~10_H \to 10_H $), which serves
the desired purpose by allowing all the desired interactions but
forbidding the dangerous $d=4$ proton decay operators and yielding
stable LSP to serve as CDM. Given the net advantages of the
minimal Higgs system $H_{\rm minimal}$, as noted above, I will
proceed to present the results of \cite{BPW} which uses this
system.\footnote{Personally I feel, however, that it would be
important to explore thoroughly the theoretical and
phenomenological consequences of both the minimal and the
large-dimensional Higgs systems involving issues such as
doublet-triplet splitting, GUT-scale threshold corrections to
gauge couplings, CP and flavor violations and proton decay. The
aim would be to look for avenues by which the two systems can be
distinguished experimentally.}

The $3\times 3$ Dirac mass matrices for the four sectors
$(u,d,l,\nu)$ proposed in Ref. \cite{BPW} were motivated in part
by the notion that flavor symmetries \cite{Hall} are responsible
for the hierarchy among the elements of these matrices (i.e., for
``33"$\gg$``23"$\sim $``32''$\gg$``22"$\gg$``12"$\gg$``11", etc.),
and in part by the group theory of SO(10)/G(224), relevant to a
minimal Higgs system. Up to minor variants \cite{FN4}, they are as
follows \footnote{A somewhat analogous scheme based on low
dimensional $SO(10)$ Higgs multiplets, has been proposed by C.
Albright and S. Barr [AB] \cite{AlbrightBarr}, who, however use
two pairs of ($16_H,~\overline{16_H}$), while BPW use only one.
One major difference between the work of AB and that of BPW
\cite{BPW} (stemming from the use of two pairs of
($16_H,~\overline{16_H}$) by AB compared to one by BPW) is that
the AB model introduces the so-called ``lop-sided'' pattern in
which some of the ``23'' and ``32'' elements are even greater than
the ``33'' element; in the BPW model on the other hand, the
pattern is consistently hierarchical with individual ``23'' and
``32'' elements (like $\eta$, $\epsilon$ and $\sigma$) being much
smaller in magnitude than the ``33'' element of 1. It turns out
that this difference leads to a characteristically different
explanation for the large (maximal) $\nu_\mu-\nu_\tau$ oscillation
angle in the two models, and in particular to a much more enhanced
rate for $\mu\to e\gamma$ in the AB model compared to that in the
BPW model (see Sec. 7).}$^,$\footnote{An alternate SO(10)-based
pattern differing from BPW and AB models is proposed in
\cite{Raby}.}:
\begin{eqnarray}
\label{eq:mat}
\begin{array}{cc}
M_u=\left[
\begin{array}{ccc}
0&\epsilon'&0\\-\epsilon'&\zeta_{22}^u&\sigma+\epsilon\\0&\sigma-\epsilon&1
\end{array}\right]{\cal M}_u^0;&
M_d=\left[
\begin{array}{ccc}
0&\eta'+\epsilon'&0\\
\eta'-\epsilon'&\zeta_{22}^d&\eta+\epsilon\\0& \eta-\epsilon&1
\end{array}\right]{\cal M}_d^0\\
&\\
M_\nu^D=\left[
\begin{array}{ccc}
0&-3\epsilon'&0\\3\epsilon'&\zeta_{22}^u&\sigma-3\epsilon\\
0&\sigma+3\epsilon&1\end{array}\right]{\cal M}_u^0;& M_l=\left[
\begin{array}{ccc}
0&\eta'-3\epsilon'&0\\
\eta'+3\epsilon'&\zeta_{22}^d&\eta-3\epsilon\\0& \eta+3\epsilon&1
\end{array}\right]{\cal M}_d^0\\
\end{array}
\end{eqnarray}
These matrices are defined in the gauge basis and are multiplied
by $\bar\Psi_L$ on left and $\Psi_R$ on right. For instance, the
row and column indices of $M_u$ are given by $(\bar u_L, \bar c_L,
\bar t_L)$ and $(u_R, c_R, t_R)$ respectively. Note the
group-theoretic up-down and quark-lepton correlations: the same
$\sigma$ occurs in $M_u$ and $M_\nu^D$, and the same $\eta$ occurs
in $M_d$ and $M_l$. It will become clear that the $\epsilon$ and
$\epsilon'$ entries are proportional to $B-L$ and are
antisymmetric in the family space (as shown above). Thus, the same
$\epsilon$ and $\epsilon'$ occur in both ($M_u$ and $M_d$) and
also in ($M_\nu^D$ and $M_l$), but $\epsilon\rightarrow
-3\epsilon$ and $\epsilon'\rightarrow -3\epsilon'$ as
$q\rightarrow l$. Such correlations result in enormous reduction
of parameters and thus in increased predictiveness. Such a pattern
for the mass-matrices can be obtained, using a minimal Higgs
system ${\bf 45}_H,{\bf 16}_H,{\bf \overline {16}}_H \mbox{ and
}{\bf 10}_H $
 and a singlet $S$ of SO(10), through effective couplings as follows \cite{FN26} (see Ref.\cite{BPW} and \cite{JCPKeK} for details):
\begin{eqnarray}
\label{eq:Yuk} {\cal L}_{\rm Yuk} &=& h_{33}{\bf 16}_3{\bf
16}_3{\bf 10}_H  + [ h_{23}{\bf 16}_2{\bf 16}_3{\bf
10}_H(S/M) \nonumber \\
&+& a_{23}{\bf 16}_2{\bf 16}_3{\bf 10}_H ({\bf
45}_H/M')(S/M)^p+g_{23}{\bf 16}_2{\bf 16}_3{\bf 16}_H^d ({\bf
16}_H/M'')(S/M)^q] \nonumber \\ &+& \left[h_{22}{\bf 16}_2{\bf
16}_2{\bf 10}_H(S/M)^2+g_{22}{\bf 16}_2{\bf 16}_2 {\bf
16}_H^d({\bf 16}_H/M'')(S/M)^{q+1} \right] \nonumber \\ &+&
\left[g_{12}{\bf 16}_1{\bf 16}_2 {\bf 16}_H^d({\bf
16}_H/M'')(S/M)^{q+2}+ a_{12}{\bf 16}_1{\bf 16}_2 {\bf 10}_H({\bf
45}_H/M')(S/M)^{p+2} \right]~.
\end{eqnarray}
Typically we expect $M'$, $M''$ and $M$ to be of order $M_{\rm
string}$ or (possibly) of order $M_{GUT}$\cite{FN6}. The VEV's of
$\langle{\bf 45}_H\rangle$ (along $B-L$), $\langle{\bf
16}_H\rangle=\langle{\bf\overline {16}}_H\rangle$ (along standard
model singlet sneutrino-like component) and of the SO(10)-singlet
$\langle S \rangle$ are of the GUT-scale, while those of ${\bf
10}_H$ and of the down type SU(2)$_L$-doublet component in ${\bf
16}_H$ (denoted by ${\bf 16}_H^d$) are of the electroweak scale
\cite{BPW,FN7}. Depending upon whether $M'(M'')\sim M_{\rm GUT}$
or $M_{\rm string}$ (see \cite{FN6}), the exponent $p(q)$ is
either one or zero \cite{FN8}.

The entries 1 and $\sigma$ arise respectively from $h_{33}$ and
$h_{23}$ couplings, while $\hat\eta\equiv\eta-\sigma$ and $\eta'$
arise respectively from $g_{23}$ and $g_{12}$-couplings. The
$(B-L)$-dependent antisymmetric entries $\epsilon$ and $\epsilon'$
arise respectively from the $a_{23}$ and $a_{12}$ couplings.
[Effectively, with $\langle{\bf 45}_H\rangle\propto$ $B-L$, the
product ${\bf 10}_H\times{\bf 45}_H$ contributes as a {\bf 120},
whose coupling is family-antisymmetric.] The relatively small
entry $\zeta_{22}^u$ arises from the $h_{22}$-coupling, while
$\zeta_{22}^d$ arises from the joint contributions of $h_{22}$ and
$g_{22}$-couplings.

Such a hierarchical form of the mass-matrices, with $h_{33}$-term
being dominant, is attributed in part to a U(1)-flavor gauge
symmetry \cite{JCPKeK,BPR} that distinguishes between the three
families and introduces powers of $\langle S \rangle/M\sim 1/10$,
and in part to higher dimensional operators involving for example
$\langle{\bf 45}_H\rangle/M'$ or $\langle{\bf 16}_H\rangle/M''$,
which are suppressed by $M_{\rm GUT}/M_{\rm string}\sim 1/10$, if
$M'$ and/or $M''\sim M_{\rm string}$.

The right-handed neutrino masses arise from the effective
couplings of the form \cite{FN30}:
\begin{equation}
\label{eq:LMaj} \mathcal{L}_{\mathrm{Maj}} = f_{ij} \mathbf{16}_i
\mathbf{16}_j \overline{\mathbf{16}}_H \overline{\mathbf{16}}_H/M
\end{equation}

\noindent where the $f_{ij}$'s include appropriate powers of
$\langle S \rangle/M$. The hierarchical form of the Majorana
mass-matrix for the RH neutrinos is \cite{BPW}:
\begin{eqnarray}
\label{eq:MajMM} M_R^\nu=\left[
\begin{array}{ccc}
x & 0 & z \\
0 & 0 & y \\
z & y & 1
\end{array}
\right]M_R
\end{eqnarray}

Following flavor charge assignments (see \cite{JCPKeK}), we have
$1\gg y \gg z \gg x$. The magnitude of M$_{\rm R}$ is estimated by
putting $f_{33}\approx 1$ and $\langle
\overline{\mathbf{16}}_H\rangle\approx M_{GUT}\approx 2\times
10^{16}$ GeV. We expect that the effective scale M of Eq.
(\ref{eq:LMaj}) should lie between $M_{string}\approx 4\times
10^{17}$ GeV and $(M_{Pl})_{reduced}\approx 2\times 10^{18}$ GeV.
Thus we take $M\approx 10^{18}$ GeV (1/2--2) \cite{BPW,JCPKeK}. We
then get the Majorana mass of the heaviest RH neutrino to be given
by $M_3\approx M_R = f_{33} \langle \overline{\mathbf{16}}_H
\rangle^2/M \approx (4\times 10^{14}\mbox{ GeV})(1/2\mbox{--}2)$.

Ignoring possible phases in the parameters and thus the source of
CP violation for a moment, and also setting $\zeta_{22}^d =
\zeta_{22}^u = 0$, as was done in Ref. \cite{BPW}, the parameters
$(\sigma,\eta, \epsilon, \epsilon',\eta', {\cal M}_u^0,\ {\rm
and}\ {\cal M}_d^0)$ can be determined by using, for example,
$m_t^{\rm phys}=174$ GeV, $m_c(m_c)=1.37$ GeV, $m_s(1\mbox{
GeV})=110-116$ MeV, $m_u(1\mbox{ GeV})=6$ MeV, and the observed
masses of $e$, $\mu$, and $\tau$ as inputs. One is thus led, {\it
for this CP conserving case}, to the following fit for the
parameters, and the associated predictions \cite{BPW}:
\begin{eqnarray}
\label{eq:fit}
\begin{array}{l}
\ \sigma\approx 0.110, \quad \eta\approx 0.151, \quad
\epsilon\approx -0.095,
 \quad |\eta'|\approx 4.4 \times 10^{-3},\\
\begin{array}{l}
\epsilon'\approx 2\times 10^{-4},\quad {\cal M}_u^0\approx
m_t(M_X)\approx 100 \mbox{ GeV},\quad {\cal M}^0_d\approx
m_{\tau}(M_X)\approx 1.1 \mbox{ GeV}.
\end{array}
\end{array}
\end{eqnarray}
These output parameters remain stable to within 10\% corresponding
to small variations ($\lsim 10$\%) in the input parameters of
$m_{t}$, $m_{c}$, $m_{s}$, and $m_{u}$. These in turn lead to the
following predictions for the quarks and light neutrinos
\cite{BPW}, \cite{JCPKeK}:
\begin{eqnarray}
\label{eq:pred}
\begin{array}{l}
m_b(m_b) \approx (4.7\mbox{--}4.9) \mbox{\ GeV},\\
\sqrt{\Delta m_{23}^2} \approx m(\nu_3) \approx \mbox{(1/24 eV)(1/2--2)},\\
\begin{array}{lcl}
V_{cb} & \approx &
\left|\sqrt{\frac{m_s}{m_b}\left|\frac{\eta+\epsilon}
{\eta-\epsilon}\right|} - \sqrt{\frac{m_c}{m_t}\left|\frac{\sigma
+\epsilon}{\sigma-\epsilon}\right|}\right| \\
 & \approx & 0.044,
\end{array}\\
\left\{ \begin{array}{lcl}
\theta^{\mathrm{osc}}_{\nu_{\mu}\nu_{\tau}} & \approx &
\left|\sqrt{\frac{m_\mu}{m_\tau}} \left|
\frac{\eta-3\epsilon}{\eta+3\epsilon} \right|^{1/2} +
\sqrt{\frac{m_{\nu_2}}{m_{\nu_3}}}\right| \\
& \approx & |0.437+(0.378\pm 0.03)| {\mbox{ (for $\frac{m(\nu_2)}{m(\nu_3)}\approx 1/6$),}}\\
\multicolumn{3}{l}{\mbox{Thus, } \sin^2
2\theta^{\mathrm{osc}}_{\nu_{\mu}\nu_{\tau}}\approx 0.993,}
\end{array}\right.\\
V_{us}\approx
\left|\sqrt{\frac{m_d}{m_s}}-\sqrt{\frac{m_u}{m_c}}\right|
\approx 0.20,\\
\left|\frac{V_{ub}}{V_{cb}} \right|\approx
\sqrt{\frac{m_u}{m_c}}\approx
0.07,\\
m_d(\mbox{1 GeV})\approx \mbox{8 MeV}.
\end{array}
\end{eqnarray}

It has been noted \cite{JCPErice,JCPKeK} that small non-seesaw
contribution to $\nu_L^e\nu_L^{\mu}$ mass term ($\sim $ few
$\times 10^{-3}$ eV) which can arise through higher dimensional
operators in accord with flavor symmetry, but which have been
ignored in the analysis given above, can lead quite plausibly to
large $\nu_e-\nu_{\mu}$ oscillation angle in accord with the LMA
MSW solution for the solar neutrino problem. Including the seesaw
contribution obtained by combining $M_\nu^D$ (Eq. (\ref{eq:mat}))
and $M^\nu_R$ (Eq. (\ref{eq:MajMM})) and with an input value of
$y\approx -1/17$ (Note that by flavor symmetry \cite{JCPKeK}, we
{\it a priori} expect $|y|\sim 1/10$) we get:
\begin{eqnarray}
\label{eq:nu12}
\begin{array}{l}
\ \ \ m(\nu_2)\approx (6-7)\times 10^{-3} \quad {\rm eV\ (from\ seesaw)}~~~~~~~~~~~~~~~~~~~~~~~~~~~~~~~~~~~~~~~(a)\\
\begin{array}{l}
\ \ m(\nu_1)\approx (1-few)\times 10^{-3};\quad {\rm thus\ \Delta
m^2_{12}\approx (3-5)\times 10^{-5} eV^2} ~~~~~~~~~~~~~~~ (b)\\
\begin{array}{l}
\ \sin^2 2\theta^{osc}_{\nu_e \nu_\mu}\approx (0.5-0.7)\quad {\rm
(from\ non\ seesaw)}~~~~~~~~~~~~~~~~~~~~~~~~~~~~~~~~~~~~~ (c)\\
\begin{array}{l}
\theta_{13}\lsim (2-5)\times 10^{-2}\quad ~~~~~~~~~~~~~~~~~~~~~~~~~~~~~~~~~~~~~~~~~~~~~~~~~~~~~~~~~~~~~~~~~(d)\\
\end{array}
\end{array}
\end{array}
\end{array}
\end{eqnarray}

While the results in Eq. (\ref{eq:pred}) are compelling
predictions of the model, the LMA-compatible solution for
$\theta^{osc}_{\nu_e \nu_\mu}$ listed in (\ref{eq:nu12})(c) should
be regarded as a plausible and consistent possibility rather than
as a compelling prediction of the framework.

The Majorana masses of the RH neutrinos ($N_{iR}\equiv N_i$) are
given by \cite{JCPErice}:
\begin{eqnarray}
\label{eq:MajM}
M_{3}& \approx & M_R\approx 4\times10^{14}\mbox{ GeV (1/2-2)},\nonumber\\
M_{2}& \approx & |y^2|M_{3}\approx \mbox{$10^{12}$ GeV(1/2-2)},\\
M_{1}& \approx & |x-z^2|M_{3} \sim (1/4\mbox{-}2)10^{-4}M_{3} \nonumber \\
 & & \sim 4\times 10^{10} \mbox{\ GeV}(1/8-4).\nonumber
\end{eqnarray}

\noindent where $y\approx -1/17$ and $x\sim z^2\sim
10^{-4}(1/2-2)$ have been used, in accord with flavor-symmetry
\cite{JCPKeK}. {\it Note that we necessarily have a hierarchical
pattern for the light as well as the heavy neutrinos with normal
hierarchy $m_1\lsim m_2\ll m_3$ and $M_1\ll M_2 \ll M_3$.}

Leaving aside therefore the question of the $\nu_e-\nu_{\mu}$
oscillation angle, it seems quite remarkable that all seven
predictions in Eq.(\ref{eq:pred}) agree with observations to
within 10\%. Particularly intriguing is the $(B-L$)-dependent
group-theoretic correlation between $V_{cb}$ and
$\theta_{\nu_{\mu}\nu_{\tau}}^{osc}$, which explains
simultaneously why one is small ($V_{cb}$) and the other is so
large ($\theta_{\nu_{\mu}\nu_{\tau}}^{osc}$) \cite{BPW,JCPKeK}.

\noindent {\bf Why $V_{cb}$ is small while $\theta^{osc}_{\nu_\mu
\nu_\tau}$ is large?}

A Comment is in order about this last feature. Often it has been
remarked by several authors that while the ``observed'' near
equality of $m_b$ and $m_\tau$ at the GUT-scale supports
quark-lepton unification, the sharp difference between $V_{cb}$
versus $\theta^{osc}_{\nu_\mu \nu_\tau}$ disfavors such a
unification. I believe that the truth is quite the opposite. This
becomes apparent if one notices a simple group-theoretic property
{\it of the minimal Higgs system $\left( \mathbf{45_{H}},
\mathbf{16_{H}}, \overline{\mathbf{16}}_{\mathbf{H}},
\mathbf{10_{H}} \right)$.} While such a system makes SU(4)-color
preserving family-symmetric contributions to fermion masses
through $\langle\mathbf{10_{H}}\rangle$ (which yields $m_b^\circ =
m_\tau^\circ$), {\it it can make SU(4)-color breaking ($B-L$)-
dependent contribution denoted by ``$\epsilon$'' (see Eq.
(\ref{eq:mat})) only through the combination
$\langle\mathbf{10_{H}}\rangle. \langle\mathbf{45_{H}}\rangle/M$,
which, however, is family-antisymmetric.} As a result, the
$(B-L)$-dependent contribution enters into the ``23'' and the
``32'' entries but not into the ``33''-entry (see Eq.
(\ref{eq:mat})).

With ``$\epsilon$'' being hierarchical (of order 1/10), following
diagonalization, this in turn means that the SU(4)-color breaking
effect for the masses of the third family-fermions are small (of
order $\epsilon^2$) as desired to preserve the near equality
$m_b^\circ \approx m_\tau^\circ$; but such breaking effects are
necessarily large for the masses of the second family fermions
(likewise for the first family), again just as desired to account
for $m_\mu^\circ \ne m_s^\circ$. The SU(4)-color breaking effects
are also large for the mixings between second and the third family
fermions (arising from the ``23'' and ``32'' entries), which
precisely explain why $V_{cb}\ll \theta^{osc}_{\nu_\mu \nu_\tau}$
and yet $m_b^\circ \approx m_\tau^\circ$.

To be specific, it may be noted from the expressions for $V_{cb}$
and $\theta^{osc}_{\nu_\mu \nu_\tau}$ in Eq. (\ref{eq:pred}), that
while the family asymmetric and ($B-L$)- dependent square root
factors like $(\eta+\epsilon/\eta-\epsilon)^{1/2}$ suppress
$V_{cb}$, if $\epsilon$ is relatively negative compared to $\eta$,
the analogous factor $(\eta-3\epsilon/\eta+3\epsilon)^{1/2}$,
necessarily enhances $\theta^{osc}_{\nu_\mu \nu_\tau}$ in a
predictable manner for the same sign of $\epsilon$ relative to
$\eta$ (the magnitudes of $\eta, \sigma$ and $\epsilon$ are of
course fixed by quark-lepton masses \cite{BPW}). In other words,
this correlation between the suppression of $V_{cb}$ and the
enhancement of $\theta^{osc}_{\nu_\mu \nu_\tau}$ has come about
due to the group theoretic property of
$\langle\mathbf{10_{H}}\rangle. \langle\mathbf{45_{H}}\rangle/M$
which is proportional to $B-L$, but family-antisymmetric. Note
this correlation would be absent if $126_H$ were used to introduce
$(B-L)-$dependence because its contributions would be
family-symmetric, and the corresponding square root factors would
reduce to unity.

Another interesting point of the hierarchical BPW model is that
with $|y|$ being hierarchical (of order 1/10 as opposed to being
of order 1) and $m(\nu_2)/m(\nu_3)$ being of order 1/5--1/10, it
as shown in Ref. \cite{BPW} that {\it the mixing angle from the
neutrino sector $\sqrt{m(\nu_2)/m(\nu_3)}$ necessarily add (rather
than subtract) to the contribution from the charged lepton sector
(see Eq. (\ref{eq:pred}))}. As a result, in the BPW model, both
charged lepton and neutrino-sectors give medium-large contribution
($\approx 0.4$) which add to naturally yield a maximal
$\theta^{osc}_{\nu_\mu \nu_\tau}$. This thus becomes a simple and
compelling prediction of the model, based essentially on the group
theory of the minimal Higgs system in the context of SO(10) or
G(224) and the hierarchical nature of the
mass-matrices.\footnote{The explanation of the largeness of
$\theta^{osc}_{\nu_\mu \nu_\tau}$ together with the smallness of
$V_{cb}$ outlined above, based on medium-large contributions from
the charged lepton and neutrino sectors, is quite distinct from
alternative explanations. In paricular, in the lop-sided
Albright-Barr model \cite{AlbrightBarr}, the largeness of
$\theta^{osc}_{\nu_\mu \nu_\tau}$ arises almost entirely from the
lop-sidedness of the charged lepton mass matrix. This distinction
between the BPW and the AB models leads to markedly different
predictions for the rate of $\mu\to e\gamma$ decay in the two
models (see remarks later).}.

The success of the model as regards the seven predictions listed
above provides some confidence in the ${\it gross\ pattern}$ of
the Dirac mass matrices presented above and motivates the study of
CP and flavor violations and baryogenesis within the same
framework. This is what I do in the next sections.

\section{CP and Flavor Violations in the SUSY SO(10)/G(224)
Framework}
\subsection{Some Experimental Facts}
On the experimental side there are now four well measured
quantities reflecting CP and/or $\Delta F = 2$ flavor violations.
They are:$\footnote{$\epsilon'_K$ reflecting direct $\Delta F = 1$
CP violation is well measured, but its theoretical implications
are at present unclear due to uncertainties in the matrix element.
We discuss this later. }$
\begin{eqnarray}
\label{eq:4qtty} \Delta m_K,\  \epsilon_K,\  \Delta m_{B_d}\ {\rm
and} \ S(B_d \rightarrow J/\Psi K_S)
\end{eqnarray}
where $S(B_d \rightarrow J/\Psi K_S)$ denotes the asymmetry
parameter in ($B_d$ versus $\overline{{B_d}})\rightarrow J/\Psi
K_S$ decays. It is indeed remarkable that the observed values
including the signs of all four quantities as well as the
empirical lower limit on $\Delta m_{B_s}$ can consistently be
realized within the standard CKM model for a single choice of the
Wolfenstein parameters \cite{Ciuchinietal}:
\begin{eqnarray}
\label{eq:rhoeta} \bar{\rho}_W\ = 0.178\ \pm\ 0.046;\
\bar{\eta}_W\ = 0.341\ \pm\ 0.028~.
\end{eqnarray}
This fit is obtained using the observed values of $\epsilon_K$ =
2.27$\times10^{-3}$, $V_{us}$ = 0.2240 $\pm$ 0.0036, $|V_{ub}|$ =
(3.30 $\pm$ 0.24)$\times 10^{-3}$, $|V_{cb}|$
 = (4.14 $\pm$
0.07)$\times 10^{-2}$ , $| \Delta m_{B_d}|$ = (3.3 $\pm$ 0.06)
$\times 10^{-13}$ GeV and $\Delta m_{B_d}/\Delta m_{B_s}$ $
>$ 0.035, and allowing
for uncertainties in the hadronic matrix elements of up to 15\%.
One can then predict the asymmetry parameter $S(B_d \rightarrow
J/\Psi K_S)$ in the SM to be $\approx 0.685 \pm 0.052
$\cite{Ciuchinietal, ASoni}. This agrees remarkably well with the
observed value $S(B_d \rightarrow J/\Psi K_S)_{expt.}$ = $0.734
\pm 0.054$, representing an average of the BABAR and BELLE results
\cite{BabarBelle}. This agreement of the SM prediction with the
data in turn poses a challenge for physics beyond the SM,
especially for supersymmetric grand unified (SUSY GUT) models, as
these generically possess new sources of CP and flavor violations
beyond those of the SM.

\subsection{Origin of CKM CP Violation in SO(10)/G(224)}
At the outset I need to say a few words about the origin of CP
violation within the G(224)/SO(10)-framework presented above.
Following Ref. \cite{BPW}, the discussion so far has ignored, for
the sake of simplicity, possible CP violating phases in the
parameters ($\sigma$, $\eta$, $\epsilon$, $\eta'$, $\epsilon'$,
$\zeta_{22}^{u,d}$, $y$, $z$, and $x$) of the Dirac and Majorana
mass matrices [Eqs.~(\ref{eq:mat}, and (\ref{eq:MajMM})]. In
general, however, these parameters can and generically will have
phases \cite{FN34}. Some combinations of these phases enter into
the CKM matrix and define the Wolfenstein parameters $\rho_W$ and
$\eta_W$, which in turn induce CP violation by utilizing the
standard model interactions.It should be stressed, however, {\it
that the values of $(\bar\rho_W, ~\bar\eta_W)$ obtained this way
from a given pattern of mass matrices based on SO(10) (as in Eq.
(\ref{eq:mat})) need not agree (even nearly) with the SM-based
phenomenological values shown in Eq. (\ref{eq:rhoeta}), for any
choice of phases of the parameters of the mass-matrices}. That in
turn would pose a challenge for the SO(10)-model in question as to
whether it can adequately describe observed CP and flavor
violations (see discussion below).

We choose to diagonalize the quark mass matrices $M_u$ and $M_d$
at the GUT scale $\sim 2\times 10^{16}$ GeV, by bi-unitary
transformations - i.e.
 \begin{eqnarray}
 \label{eq:xdxu}
M_d^{diag}\ =\ X_L^{d\dagger}M_d X_R^{d}\ {\rm and} \ M_u^{diag}\
=\ X_L^{u\dagger}M_u X_R^{u}
\end{eqnarray}
with phases of $q_{L,R}^i$ chosen such that the eigenvalues are
real and positive and that the CKM matrix $V_{CKM}$ (defined
below) has the Wolfenstein form \cite{Wolfenstein}). Approximate
analytic expressions for $X_{L,R}$ are given in Ref. \cite{BPR}.

The CKM elements in the Wolfenstein basis are given by the matrix
$V_{CKM}=e^{-i\alpha}(X_L^{u \dagger} X_L^d)$, where
$\alpha=(\phi_{\sigma-\epsilon}-\phi_{\eta-\epsilon})-(\phi_{\epsilon'}-\phi_{\eta'+\epsilon'})$.

\subsection{SUSY CP and Flavor Violations}
{\bf SUSY Breaking}

As is well known, since the model is supersymmetric, non-standard
CP and flavor violations would generically arise in the model
through sfermion/gaugino quantum loops involving scalar $(mass)^2$
transitions \cite{HKR}. The latter can either preserve chirality
(as in $\tilde{q}^i_{L,R}\rightarrow \tilde{q}^j_{L,R}$) or flip
chirality (as in $\tilde{q}^i_{L,R}\rightarrow
\tilde{q}^j_{R,L}$). Subject to our assumption on SUSY breaking
(specified below), it would turn out that these scalar $(mass)^2$
parameters get completely determined within our model by the
fermion mass-matrices, and the few parameters of SUSY breaking.

We assume that flavor-universal soft SUSY-breaking is transmitted
to the SM-sector at a messenger scale M$^*$, where M$_{GUT}< $
M$^{*}\le$ M$_{string}$. This may naturally be realized e.g. in
models of mSUGRA \cite{msugra}, or gaugino-mediation
\cite{gauginomed}. With the assumption of extreme universality as
in CMSSM, supersymmetry introduces five parameters at the scale
M$^{*}$:
\begin{center}
$m_o, m_{1/2}, A_o, \tan\beta\ {\rm and}\ sgn(\mu).$
\end{center}

\noindent For most purposes, we will adopt this restricted version
of SUSY breaking with the added restriction that $A_o$ = 0 at
M$^{*}$ \cite{gauginomed}. However, we will not insist on strict
Higgs-squark-slepton mass universality. Even though we have flavor
preservation at M$^{*}$, flavor violating scalar
(mass)$^2$--transitions and A-terms arise in the model through RG
running from M$^*$ to $M_{GUT}$ and from $M_{GUT}$ to the EW
scale. As described below, we thereby have {\it three sources} of
flavor violation.

\noindent {\bf (i) RG Running of Scalar Masses from M$^{*}$ to
M$_{\rm GUT}$.}

With family universality at the scale M$^{*}$, all sfermions have
the mass m$_o$ at this scale and the scalar (mass)$^2$ matrices
are diagonal. Due to flavor dependent Yukawa couplings, with $h_t
= h_b = h_\tau (= h_{33})$ being the largest, RG running from
M$^{*}$ to M$_{\rm GUT}$ renders the third family lighter than the
first two (see e.g. \cite{BarbieriHallStrumia}) by the amount:
\begin{eqnarray}
\label{eq:deltambr} \Delta\hat{m}_{\tilde{b}_{L}}^2 =
\Delta\hat{m}_{\tilde{b}_{R}}^2 =
\Delta\hat{m}_{\tilde{\tau}_{L}}^2 =
\Delta\hat{m}_{\tilde{\tau}_{R}}^2 \equiv\Delta\approx
-\bigl(\frac{30m_o^2}{16\pi^2}\bigr) h_t^2\ ln(M^{*}/M_{GUT})~.
\end{eqnarray}
Note the large coefficient ``30'', which is a consequence of
SO(10). The factor 30$\to$12 for the case of G(224). The squark
and slepton (mass)$^2$ matrices thus have the form $\tilde{\rm
M}^{(o)}$ = diag(m$_o^2$, m$_o^2$, m$_o^2 -\Delta$). Transforming
$\tilde{\rm M}^{(o)}$ by $X_{L,R}^f$, which diagonalize fermion
mass-matrices, i.e. evaluating X$_L^{f\dagger}$($\tilde{\rm
M}^{(o)}$)$_{LL}$ X$_L^f$ and similarly for L$\to$R, where f = u,
d, l, introduces off-diagonal elements in the so-called SUSY basis
(at the GUT-scale) given by:
\begin{eqnarray}
\label{eq:deltahat} (\hat{\delta}_{LL,RR}^f)_{ij}=
\left(X_{L,R}^{f\dagger}(\tilde{\rm M}^{(o)})
X_{L,R}^f\right)_{ij}/m^2_{\tilde{f}}
\end{eqnarray}
\noindent These induce flavor and CP violating transitions
$\tilde{q}_{L,R}^i\to \tilde{q}_{L,R}^j$ and $\tilde{l}_{L,R}^i\to
\tilde{l}_{L,R}^j$. Note that these transitions depend upon the
matrices $X_{L,R}^f$, which are of course determined by the
entries (including phases) in the fermion mass matrices (Eq.
(\ref{eq:mat})). Here m$_{\tilde{f}}$ denotes an average squark or
slepton mass (as appropriate) and the hat signifies GUT-scale
values.

\noindent {\bf (ii) RG Running of the $A-$parameters from M$^{*}$
to M$_{\rm GUT}$.}

Even if  $A_o$ = 0 at the scale M$^{*}$ (as we assume for
concreteness, see also \cite{gauginomed}). RG running from M$^{*}$
to M$_{\rm GUT}$ induces $A-$parameters at M$_{\rm GUT}$,
involving the SO(10)/G(224) gauginos and yukawa couplings
\cite{BarbieriHallStrumia}; these yield chirality flipping
transitions $\tilde{q}_{L,R}^i\to \tilde{q}_{R,L}^j$ and
($\tilde{l}^i_{L,R}\to \tilde{l}^j_{R,L}$). Because of large
SO(10) Casimirs, these induced A-terms arising from post-GUT
physics are large even if $\ln(M^*/M_{GUT})\approx 1$. The
chirality flipping transition angles are given by:
\begin{eqnarray}
\label{eq:deltalr} (\delta^{f}_{LR})_{ij}\ \equiv\
(A^{f}_{LR})_{ij}\ \bigl(\frac{v_f}{m_{\tilde{f}}^2}\bigr).
\end{eqnarray}

Here f = u, d, l. The matrices $A_{LR}^{ij}$ are given explicitly
in Refs. \cite{BPR} and \cite{LFV}. Note that these induced
A-terms are also completely determined by the fermion mass
matrices, for any given choice of the universal SUSY parameters
($m_o,~m_{1/2},~\tan\beta$ and $M^*$).

\noindent $\bf{ (iii)\ Flavor\ Violation\ Through\ RG\ Running\
From\ M_{GUT}\ to\ m_W\ in\ MSSM}:$ It is well known that, even
with universal masses at the GUT scale, RG running from $M_{GUT}$
to $m_W$ in MSSM, involving contribution from the top Yukawa
coupling, gives a significant correction to the mass of
$\tilde{b}'_L =
V_{td}\tilde{d}_L+V_{ts}\tilde{s}_L+V_{tb}\tilde{b}_L$, which is
not shared by the mass-shifts of $\tilde{b}_R, \tilde{d}_{L,R}$
and $\tilde{s}_{L,R}$. This in turn induces flavor violation.
Here, $ \tilde{d}_L, \tilde{s}_L$ and $\tilde{b}_L$ are the SUSY
partners of the physical $d_L, s_L$ and $b_L$ respectively. The
differential mass shift of $\tilde{b}'_L$ arising as above, may be
expressed by an effective Lagrangian \cite{carena}: $\Delta{\cal
L} = -(\Delta m_L^{'2})\tilde{b}_L^{'*}\tilde{b}'_L$, where
\begin{eqnarray}
\label{eq:deltam'l} \Delta m_L^{'2}=-3/2 m_o^2\eta_t + 2.3 A_o
m_{1/2} \eta_t(1-\eta_t) - (A_o^2/2)\eta_t(1-\eta_t) +
m_{1/2}^2(3\eta_t^2-7\eta_t)~.
\end{eqnarray}

Here $\eta_t = (h_t/h_f)\approx (m_t/v \sin\beta)^2(1/1.21)\approx
0.836$ for tan$\beta$ = 3. Expressing $\tilde{b}'_L$ in terms of
down-flavor squarks in the SUSY basis as above, Eq.
(\ref{eq:deltam'l}) yields new contributions to off diagonal
squark mixing. Normalizing to $m_{sq}^2$, they are given by
\begin{eqnarray}
\label{eq:delta'll} \delta_{LL}^{'(12,13,23)} =\bigl(\frac{\Delta
m_L^{'2}}{m_{sq}^2}\bigr)(V_{td}^{*}V_{ts},\ V_{td}^{*}V_{tb},\
V_{ts}^{*}V_{tb})~.
\end{eqnarray}

The net chirality preserving squark $(mass)^2$ off-diagonal
elements at $m_W$ are then obtained by adding the respective
GUT-scale contributions from Eqs. (\ref{eq:deltahat}) to that from
Eq. (\ref{eq:delta'll}). They are:
\begin{eqnarray}
\label{eq:deltallrr}
\delta_{LL}^{ij}=\hat\delta_{LL}^{ij}+\delta_{LL}^{'ij};\quad\
 \delta_{RR}^{ij}=\hat\delta_{RR}^{ij}
\end{eqnarray}

\subsection{The Challenge for SUSY SO(10)/G(224)}
The interesting point is that the net values including phases of
the off-diagonal squark-mixings, arising from the three sources
listed above, and thereby the flavor and CP violations induced by
them, are entirely determined within our approach by the entries
in the quark mass-matrices and the choice of the universal SUSY
parameters ($m_0$, $m_{1/2}$, $M^*$, $\tan{\beta}$ and
sgn$(\mu)$). Within the $G(224)/SO(10)$ framework presented in
Sec.~4, the quark mass-matrices are however tightly constrained by
our considerations of fermion masses and neutrino-oscillations.

The question thus arises: {\it Can observed CP and/or
flavor-violations in the quark and lepton sectors (including the
empirical limits on some of these) emerge consistently within the
$G(224)/SO(10)$-framework, for \emph{any} choice of phases in the
fermion mass-matrices of Eq.~(\ref{eq:mat}), while preserving all
its successes with respect to fermion masses and neutrino
oscillations?}

This is indeed a \emph{non-trivial challenge} to meet within the
$SO(10)$ or $G(224)$-framework, since  the constraints from both
CP and flavor violations on the one hand and fermion masses and
neutrino oscillations on the other hand are severe.

To be specific, the fact that all four entities ($\Delta m_K,\
\epsilon_K,\ \Delta m_{B_d}$ and $S(B_d\rightarrow J/\psi K_S)$)
can be realized consistently in accord with experiments within the
standard CKM model for a single choice of the Wolfenstein
parameters $\bar\rho_W$ and $\bar\eta_W$ (Eq. (\ref{eq:rhoeta}))
strongly suggests {\it that even for the SUSY SO(10)/G(224)-model,
the corresponding SM-contributions, at least to these four
entities, should be the dominant ones, with SUSY contributions
being sub-dominant or small}.\footnote{The alternative of
SUSY-contributions being relatively important compared to the
SO(10)-based SM contributions and correcting for its pitfalls in
just the right way for each of these four entities appear to be
rather contrived and may require arbitrary adjustment of the many
MSSM parameters. Such a scenario would at the very least mean that
the good agreement between the SM-predictions and experiments is
fortuitous.} This in turn means that there should exist a choice
of the parameters of the SO(10)-based mass matrices (like
$\sigma,\ \eta,\ \epsilon,\ \epsilon'$ etc.), viewed in general as
complex, for which not only (a) the fermion masses and (b) the CKM
mixings $|V_{ij}|$ should be described correctly (as in
Eq.(\ref{eq:pred})), but also (c) the Wolfenstein parameters
$\bar\rho'_W$ and  $\bar\eta'_W$ derived from the SO(10)-based
mass-matrices should be close to the phenomenological SM values
(Eq. (\ref{eq:rhoeta})). {\it A priori}, a given SO(10)-model,
with a specified pattern for fermion mass matrices, may not in
fact be able to satisfy all three constraints (a), (b) and (c)
simultaneously.\footnote{For a discussion of the difficulties in
this regard within a recently proposed SO(10)-model see e.g. Ref.
\cite{GohMohapatNg}.}

\subsection{The Results}

Without further elaboration, I will now briefly summarize the main
results of Refs.~\cite{BPR} and \cite{LFV}.

(1) Allowing for phases ($\sim 1/10$ to $\sim 1/2$) in the
parameters $\eta$, $\sigma$, $\epsilon'$ and $\zeta^d_{22}$ of the
$G(224)/SO(10)$-framework (see Eq.~(\ref{eq:mat})) we found that
there do exist solutions which yield masses and mixings of quarks
and leptons including neutrinos, all in good accord with
observations (to within 10 \%), and at the same time yield the
following values for the Wolfenstein parameters (see Ref.
\cite{BPR} for details):
\begin{equation}
\bar\rho'_W\approx 0.15,\; \bar\eta'_W \approx 0.37.\quad({\rm
SO(10)/G(224)-model}) \label{eq:22}
\end{equation}
The prime here signifies that these are the values of $\bar\rho_W$
and $\bar\eta_W$ which are derived (for a suitable choice of
phases in the parameters of the fermion mass matrices) from within
the structure of the SO(10)-based mass-matrices
(Eq.~(\ref{eq:mat})). The corresponding phenomenological values
are listed in Eq.~(\ref{eq:rhoeta}). Note, as desired, the
$G(224)/SO(10)$-framework presented here has turned out to be
capable of yielding $\bar\rho'_W$ and $\bar\eta'_W$ close to the
SM-values of $\bar\rho_W$ and $\bar\eta_W$ while preserving the
successes with respect to fermion masses and neutrino oscillations
as in Sec.~4. As mentioned above, this is indeed a non-trivial but
most desirable feature.

(2) Including both the SM-contribution (with $\bar\rho'_W$ and
$\bar\eta'_W$ as above) and the SUSY-contribution (with a
plausible choice of the spectrum-e.g. $m_{sq}\approx (0.8-1)$ TeV
and $x=(m^2_{\tilde{g}}/m^2_{sq})\approx 0.6-0.8$), we obtain
\cite{BPR}:
\begin{eqnarray}
(\Delta m_K)_{short dist}\approx 3\times 10^{-15}\mbox{ GeV};\nonumber\\
\epsilon_K \approx(2\, \mbox{to}\, 2.5)\times 10^{-3};\nonumber\\
\Delta m_{B_d} \approx (3.5\, \mbox{to}\, 3.6) \times
10^{-13}\mbox{ GeV};
\nonumber\\
S(B_d \rightarrow J/\Psi K_s) \approx 0.68 - 0.74. \label{eq:23}
\end{eqnarray}

We have used $\hat{B}_K=0.86$ and $f_{Bd}\sqrt{\hat{B}_{Bd}}=215$
MeV (see \cite{Ciuchinietal}). Now all four on which there is
reliable data are in good agreement with observations (within
10\%). The spectrum of ($m_{sq}$, $m_{\tilde{g}}$) considered
above can be realized, for example for a choice of $(m_0,
m_{1/2})\approx (600, 220)$ GeV. For a more complete presentation
of the results involving other choices of $(m_o,~m_{1/2})$, and a
discussion on the issue of consistency with WMAP results on the
LSP as cold dark matter, see Refs. \cite{BPR} and \cite{LFV}.

In all these cases, the SUSY-contribution turns out to be rather
small ($\lsim 5 \%$ in amplitude), except however for
$\epsilon_K$, for which it is sizable ($\approx 20 - 30 \%$) and
has opposite sign, compared to the SM-contribution. Had the SUSY
contribution to $\epsilon_K$ been positive relative to the
SM-contribution, $\epsilon_K$(total) would have been too large
($\approx(3.1-3.5)\times 10^{-3}$), in strong disagreement with
the observed value of $2.27\times 10^{-3}$, despite the
uncertainty in $\hat{B}_K$. In short, {\it the SUSY contribution
of the model to $\epsilon_K$ has just the right sign and nearly
the right magnitude to play the desired role.} This seems to be an
intriguing feature of the model.

\emph{We thus see that the SUSY $G(224)$ or $SO(10)$-framework
(remarkably enough) has met all the challenges so far in being
able to reproduce the observed features of both CP and
quark-flavor violations as well as fermion masses and
neutrino-oscillations!}

\noindent {\bf Other Predictions}

Other predictions of the model which incorporate contributions
from $\delta_{LL}^{23},~\delta_{RR}^{23},~\delta_{LR}^{23}$ and
$\delta_{RL}^{23}$, include (see Ref. \cite{BPR} for details):
\begin{eqnarray}
S(B_d\rightarrow\phi K_S)(Tot\approx SM)\approx 0.65-0.73
\end{eqnarray}
\begin{eqnarray}
\Delta m_{B_s}(Tot\approx SM)\approx 17.3\ ps^{-1}
\bigl(\frac{f_{B_s}\sqrt{\hat{B}_{B_s}}}{245 MeV}\bigr)^2~.
\end{eqnarray}
\begin{eqnarray}
A(b\to s\gamma)_{SUSY}\approx (1-5)\%~~ {\rm of} ~~ A(b\to
s\gamma)_{SM}
\end{eqnarray}
\begin{eqnarray}
Re(\epsilon'/\epsilon)_{SUSY}\approx +(8.8\times
10^{-4})(B_G/4)(5/\tan\beta)~.
\end{eqnarray}

Particularly interesting is the prediction of the model that the
asymmetry parameter $S(B_d\rightarrow\phi K_S)$ should be close to
the SM value of $\approx 0.70\pm0.10$. At present, there is
conflicting data: $S(B_d\rightarrow\phi K_S)=(+0.50\pm
0.25^{+0.07}_{-0.04})_{BaBar}; (+0.06\pm 0.33\pm 0.09)_{BELLE}$
\cite{BabarBelleNew}\footnote{At the time of completing this
manuscript, the BELLE group reported a new value of
$S(B_d\rightarrow\phi K_S)=+0.44\pm 0.27^\pm 0.05$ at the 2005
Lepton-Photon Symposium \cite{NewBelle}. This value is close to
that reported by BaBar and enhances the possibilty of the true
value being close to the SM value.}. It will thus be extremely
interesting to see both from the point of view of the present
model and the SM whether the true value of $S(B_d\rightarrow\phi
K_S)$ will turn out to be close to the SM prediction or not.

\noindent {\bf EDM's}

For a representative choice of ($m_o,\ m_{1/2}$) = (600, 300) GeV
(i.e. $m_{sq}$ = 1 TeV, $m_{\tilde{g}}$ = 900 GeV, $m_{\tilde{l}}$
= 636 GeV and $m_{\tilde{B}}$ = 120 GeV), the induced A-terms (see
Eq. (\ref{eq:deltalr})) lead to \cite{BPR}:
\begin{eqnarray}
\label{eq:edmN} (d_n)_{A_{ind}}= (1.6, 1.08)\times 10^{-26} e cm\
{\rm for}\  \tan\beta = (5, 10)~.
\end{eqnarray}
\begin{eqnarray}
\label{eq:edm} (d_e)_{A_{ind}} =\frac{1.1\times
10^{-28}}{\tan\beta}\ e cm~.
\end{eqnarray}

Given the experimental limits $d_n<6.3\times 10^{-26}$ e cm
\cite{edmneutron} and $d_e<4.3\times 10^{-27}$ e cm
\cite{edmelectron}, we see that the predictions of the model
(arising only from the induced $A$-term contributions)  especially
for the EDM of the neutron is in an extremely interesting range
suggesting that it should be discovered with an improvement of the
current limit by a factor of about 10.

\section{Lepton Flavor Violation in SUSY SO(10)/G(224)}

It has been recognized for sometime that lepton flavor violating
processes (such as $\mu\to e\gamma,~\tau\to\mu\gamma,~\mu N\to e
N~ etc.$), can provide sensitive probes into new physics beyond
the SM, especially that arising in SUSY grand-unification
\cite{HKR,BarbieriHallStrumia,Masiero}, and that too with heavy
right-handed neutrinos \cite{Masiero}. In our case these get
contributions from three sources:

(i) The slepton (mass)$^2$ elements $(\delta m^2)^{ij}_{LL}$
arising from RG-running of scalar masses from $M^*\to M_{GUT}$ in
the context of SO(10)/G(224) (see Eq. (\ref{eq:deltahat})),

(ii) The chirality flipping slepton (mass)$^2$ elements $(\delta
m^2)^{ij}_{LR}$ arising from A-terms induced  through RG-running
from $M^*\to M_{GUT}$ in the context of SO(10) or G(224) (see Sec.
5.3), and

(iii) $(\delta m'^2)^{ij}_{LL}$ arising from RG-running from
$M_{GUT}$ to the RH neutrino mass-scales $M_{R_i}$ involving
$\nu_R^i$ Dirac Yukawa couplings corresponding to
Eq.(\ref{eq:mat}) which (in the leading log approximation) yield:
\begin{eqnarray}
\label{eq:RHN} (\delta_{LL}^l)_{ij}^{RHN} =
\frac{-(3m_o^2+A_o^2)}{8\pi^2}\sum_{k=1}^{3}
(Y_N)_{ik}(Y_N^{*})_{jk}\ ln(\frac{M_{GUT}}{M_{R_k}})~.
\end{eqnarray}

\noindent Note that the masses M$_{R_i}$ of RH neutrinos are
fairly well determined within the model (see Eq. (\ref{eq:MajM})).

There is a vast literature on the subject of lepton flavor
violation (LFV). (For earlier works see
Ref.~\cite{Masiero,BarbieriHallStrumia}; and for a partial list of
references including recent works see Ref.~\cite{61MVV}). Most of
the works in the literature have focused only on the contribution
from the third source, involving the Yukawa couplings of the RH
neutrinos, which is proportional to $\tan\beta$ in the amplitude.
It turns out, however, that the contributions from the first two
sources arising from post-GUT physics (i.e. $SO(10)$-running from
$M^*$ to $M_{GUT}$) are in fact the dominant ones for
$\tan\beta\lsim1 0$, as long as $\ln(M^*/M_{GUT})\gsim 1$. We
consider the contribution from all three sources by summing the
corresponding amplitudes, and by varying ($m_0$, $m_{1/2}$,
$\tan{\beta}$ and $sgn(\mu)$)

Here I present the predictions of the model for five different
choices of $(m_o,~m_{1/2})$ with $\tan\beta=10$ or 20 and
$\ln(M^*/M_{GUT})=1$, to indicate the nature of the predictions.
(Results for a wider choice of parameters and a more detailed
discussion may be found in Ref. \cite{LFV}). We have set
$(M_{R_1},~M_{R_2},~M_{R_3})=(10^{10},~10^{12},~5\times 10^{14})$
GeV (see Eq. (\ref{eq:MajM})) and $A_o(M^*)=0$. The predicted
rates for G(224) are smaller than those for SO(10) approximately
by a factor of 4 to 6 (see comments in Sec. 5). The results for
SO(10) are presented in Table 1.

\vspace*{12pt}

\noindent
\begin{table}
\begin{tabular}{|c|c|c|c|c|}
\hline \rule[-3mm]{0mm}{8mm} ($m_o,\ m_{1/2}$)//$\tan\beta$ &
\multicolumn{2}{c|}{Br($\mu\to e\gamma$)}&\multicolumn{2}{c|} {Br($\tau\to\mu\gamma$)} \\
\hline
& $\mu>0$& $\mu<0$ & $\mu>0$ &$\mu<0$ \\

\hline

I (600, 300)//10 & 3.3$\times10^{-12}$ & 9.8$\times 10^{-12}$
         & 2.4$\times 10^{-9}$  & 3.1$\times10^{-9}$\\ \hline
II (800, 250)//10 & 2.9$\times10^{-13}$ & 1.7$\times 10^{-12}$
         & 1.9$\times 10^{-9}$  & 1.9$\times10^{-9}$  \\\hline
III (450, 300)//10 & 2.7$\times10^{-11}$ & 4.6$\times 10^{-11}$
         & 2.7$\times 10^{-9}$  & 5.6$\times10^{-9}$   \\\hline
IV (500, 250)//10 & 5.9$\times10^{-12}$ & 1.9$\times 10^{-11}$
         & 4.8$\times 10^{-9}$  & 6.4$\times10^{-9}$  \\\hline
V (100, 440)//10 & 1.02$\times10^{-8}$ & 1.02$\times 10^{-8}$
         & 8.3$\times 10^{-8}$  & 8.4$\times10^{-8}$  \\ \hline
VI (1000, 250)//10 & 1.6$\times10^{-13}$ & 5.6$\times 10^{-12}$
         & 9.5$\times 10^{-10}$  & 9.0$\times10^{-10}$  \\ \hline
VII (400, 300)//20 & 9.5$\times10^{-12}$ & 3.8$\times 10^{-11}$
         & 1.4$\times 10^{-8}$  & 1.8$\times10^{-8}$  \\\hline
\end{tabular}

\caption{Branching ratios of $l_i\to l_j\gamma$ for the SO(10)
framework with $\kappa\equiv \ln(M^*/M_{GUT})=1$; ($m_o,\
m_{1/2}$) are given in GeV, which determine $\mu$ through
radiative electroweak symmetry breaking conditions. The entries
for Br($\mu\to e\gamma$) for the case of G(224) would be reduced
by a factor $\approx 4-6$ compared to that of SO(10) (see text).}
\end{table}

The following points regarding these results are worth noting:

\noindent {\bf(1)} We find that the contribution due to the
presence of the RH neutrinos\footnote{In the context of
contributions due to the RH neutrinos alone, there exists an
important distinction (partially observed by Barr, see Ref.
\cite{SO(10)LFV}) between the hierarchical BPW form \cite{BPW} and
the lop-sided Albright-Barr (AB) form \cite{AlbrightBarr} of the
mass-matrices. The amplitude for $\mu\to e\gamma$ from this source
turns out to be proportional to the difference between the
(23)-elements of the Dirac mass-matrices of the charged leptons
and the neutrinos, with (33)-element being 1. This difference is
(see Eq. (\ref{eq:mat})) is $\eta-\sigma\approx 0.041$, {\it which
is naturally small for the hierarchical BPW model} (incidentally
it is also $V_{cb}$), while it is order one for the lop-sided AB
model. This means that the rate for $\mu\to e\gamma$ due to RH
neutrinos would be about $600$ times larger in the AB model than
the BPW model (for the same input SUSY parameters). For a
comparative study of the BPW and the AB models using processes
such as $\mu\to e\gamma$ and edm's, see forthcoming paper by P.
Rastogi \cite{ABBPW}. } is about an order of magnitude smaller,
{\it in the amplitude}, than those of the others arising from
post-GUT physics (proportional to $\hat{\delta}^{ij}_{LL},\
\delta^{ij}_{LR}$ and $\delta^{ij}_{RL}$). The latter arise from
RG running of the scalar masses and the $A-$parameters in the
context of SO(10) or G(224) from $M^*$ to M$_{GUT}$. It seems to
us that the latter, which have commonly been omitted in the
literature, should exist in any SUSY GUT model for which the
messenger scale for SUSY-breaking is high ($M^*>M_{GUT}$), as in a
mSUGRA model. {\it The inclusion of these new contributions to LFV
processes arising from post-GUT physics, that too in the context
of a predictive and realistic framework, is the distinguishing
feature of the study carried out in Ref. \cite{LFV}.}\footnote{For
the sake of comparison, should one drop the post-GUT contribution
by setting $M^*=M_{GUT}$, however, the predicted $Br(\mu\to
e\gamma)$, based on RHN contributions only, would be reduced
significantly in our model to e.g. ($4.2,~2.9,$ and $8.6$)$\times
10^{-15}$ for cases I, II and IV respectively.}

\noindent {\bf(2)} Owing to the general prominence of the new
contributions from post-GUT physics, we see from table 1 that case
V, (with low $m_o$ and high $m_{1/2}$) is clearly excluded by the
empirical limit on $\mu\to e \gamma$-rate (see Sec. 1). Case III
is also excluded, for the case of SO(10), yielding a rate that
exceeds the limit by a factor of about 2 (for
$\kappa=\ln(M^*/M_{GUT})\gsim 1$), though we note that for the
case of G(224), Case III is still perfectly compatible with the
observed limit (see remark below table 1). All the other cases (I,
II, IV, VI, and VII), with medium or moderately heavy ($\gsim$ 500
GeV) sleptons , are compatible with the empirical limit, even for
the case of SO(10). The interesting point about these predictions
of our model, however, is that $\mu\to e \gamma$ should be
discovered, even with moderately heavy sleptons ($\sim 800-1000$
GeV), both for SO(10) and G(224), with improvement in the current
limit by a factor of 10--100. Such an improvement is being planned
at the forthcoming MEG experiment at PSI.

\noindent {\bf(3)} We see from table 1 that $\tau\to\mu\gamma$
(leaving aside case V, which is excluded by the limit on $\mu\to
e\gamma$), is expected to have a branching ratio in the range of
$2\times 10^{-8}$ (Case VII) to about $(1\ {\rm or}\  2)\times
10^{-9}$ (Case VI or II). The former may be probed at BABAR and
BELLE, while the latter can be reached at the LHC or a super B
factory. The process $\tau\to e\gamma$ would, however, be
inaccessible in the foreseeable future (in the context of our
model).

\noindent {\bf(4)} {\bf The WMAP-Constraint:} Of the cases
exhibited in table 1, Case V ($m_o = 100$ GeV, $m_{1/2} = 440$
GeV) would be compatible with the WMAP-constraint on relic dark
matter density, in the context of CMSSM, assuming that the
lightest neutralino is the LSP and represents cold dark matter
(CDM), accompanying co-annihilation mechanism. (See e.g.
\cite{JEllis}). As mentioned above (see table 1), a spectrum like
Case V, with low $m_o$ and higher $m_{1/2}$, is however excluded
in our model by the empirical limit on $\mu\to e\gamma$. {\it Thus
we infer that in the context of our model CDM cannot be associated
with the co-annihilation mechanism.}

Several authors (see e.g. Refs.\cite{Olive} and \cite{Baer}),
have, however considered the possibility that Higgs-squark-slepton
mass universality need not hold even if family universality does.
In the context of such non-universal Higgs mass (NUHM) models, the
authors of Ref. \cite{Baer} show that agreement with the WMAP data
can be obtained over a wide range of mSUGRA parameters. In
particular, such agreement is obtained for ($m_\phi/m_o$) of order
unity (with either sign) for almost all the cases (I, II, III, IV,
VI and VII)\footnote{We thank A. Mustafayev and H. Baer for
private communications in this regard.}, with the LSP (neutralino)
representing CDM.\footnote{We mention in passing that there may
also be other posibilities for the CDM if we allow for either
non-universal gaugino masses, or axino or gravitino as the LSP, or
R-parity violation (with e.g. axion as the CDM). } (Here
$m_\phi\equiv sign(m^2_{H_{u,d}})\sqrt{|m^2_{H_{u,d}}|}$, see
\cite{Baer}). All these cases (including Case III for G(224)) are
of course compatible with the limit on $\mu\to e\gamma$.

\noindent {\bf(5)} {\bf Coherent $\mu - e$ conversion in nuclei:}
In our framework, $\mu-e$ conversion (i.e. $\mu^- + N \rightarrow
e^- + N$)
 will occur when the photon emitted in the virtual decay
$\mu \rightarrow e \gamma^*$ is absorbed by the nucleus (see e.g.
 \cite{marciano}). In such situations, there is a rather simple relation connecting
the $\mu-e$ conversion rate with $B(\mu \rightarrow e \gamma)$:
$B(\mu \rightarrow e \gamma)/
(\omega_{conversion}/\omega_{capture}) = R \simeq (230-400)$,
depending on the nucleus.  For example, $R$ has been calculated to
be $R \simeq 389$ for $^{27}Al$, 238 for $^{48}Ti$ and 342 for
$^{208}Pb$ in this type of models. (These numbers were computed in
\cite{marciano} for the specific model of
\cite{BarbieriHallStrumia}, but they should approximately hold for
our model as well.) With the branching ratios listed in Table 1
($\sim 10^{-11}$ to $10^{-13}$) for our model,
$\omega_{conversion}/\omega_{capture} \simeq$ (40--1) $\times
10^{-15}$. The MECO experiment at Brookhaven is expected to have a
sensitivity of $10^{-16}$ for this process, and thus will test our
model.

In summary, lepton flavor violation is studied in \cite{LFV}
within a predictive SO(10)/G(224)-framework, possessing
supersymmetry, that was proposed in Refs. \cite{BPW,BPR}. The
framework seems most realistic in that it successfully describes
five phenomena: (i) fermion masses and mixings, (ii) neutrino
oscillations, (iii) CP violation, (iv) quark flavor-violations, as
well as (v) baryogenesis via leptogenesis (see below)
\cite{PatiLepto}. LFV emerges as an important prediction of this
framework bringing no new parameters, barring the few
flavor-preserving SUSY parameters.

Our results show that -- (i) The decay $\mu\to e\gamma$ should be
seen with improvement in the current limit by a factor of 10 --
100, even if sleptons are moderately heavy ($\sim 800$ GeV, say);
(ii) for the same reason, $\mu-e$ conversion ($\mu N \to e N$)
should show in the planned MECO experiment, and (iii)
$\tau\to\mu\gamma$ may be accessible at the LHC and a super
B-factory.

\section{Baryogenesis Via Leptogenesis Within the $G(224)/SO(10)$-Framework}

The observed matter-antimatter asymmetry provides an important
clue to physics at truly short distances. Given the existence of
the RH neutrinos, as required by the symmetry $SU(4)$-color or
$SU(2)_R$, possessing superheavy Majorana masses which violate B-L
by two units, baryogenesis via leptogenesis \cite{Yanagida,FYKRS}
has emerged as perhaps the most viable and natural mechanism for
generating the baryon asymmetry of the universe. The most
interesting aspect of this mechanism is that it directly relates
our understanding of the light neutrino masses to our own origin.
The question of whether this mechanism can quantitatively explain
the magnitude of the observed baryon-asymmetry depends however
crucially on the Dirac as well as the Majorana mass-matrices of
the neutrinos, including the phases and the eigenvalues of the
latter-i.e. $M_1$, $M_2$ and $M_3$ (see Eq.~(\ref{eq:MajM})).

This question has been considered in a recent work
\cite{PatiLepto} in the context of a realistic and predictive
framework for fermion masses and neutrino oscillations, based on
the symmetry $G(224)$ or $SO(10)$ , as discussed in Sec.~4, with
CP violation treated as in Sec.~5. It has also been discussed in a
recent review \cite{JCPErice}. Here I will primarily quote the
results and refer the reader to Ref.~\cite{PatiLepto} for more
details especially including the discussion on inflation and
relevant references.

The basic picture is this. Following inflation, the lightest RH
neutrinos ($N_1$'s) with a mass $\approx 10^{10}$ GeV ($1/3\ -\
3$) are produced either from the thermal bath following reheating
($T_{RH}\approx \mbox{ few} \times 10^9$ GeV), or non-thermally
directly from the decay of the inflaton \footnote{In this case the
inflaton can naturally be composed of the Higgs-like objects
having the quantum numbers of the RH sneutrinos
($\tilde{\nu}_{RH}$ and $\tilde{\bar{\nu}}_{RH}$) lying in $(1,\
2,\ 4)_H$ and $(1,\ 2,\ \bar{4})_H$ for $G(224)$ (or $16_H$ and
$\bar{16}_H$ for $SO(10)$), whose VEV's break B-L
 and give Majorana masses to the RH neutrinos via the coupling shown in
Eq.~(\ref{eq:LMaj}).} (with $T_{RH}$ in this case being about
$10^7\ -\ 10^8$ GeV). In either case, the RH neutrinos having
Majorana masses decay by utilizing their Dirac Yukawa couplings
into both $l+H$ and $\bar{l}+\bar{H}$ (and corresponding SUSY
modes), thus violating B-L. In the presence of CP violating
phases, these decays produce a net lepton-asymmetry
$Y_L=(n_L-n_{\bar{L}})/s$ which is converted to a baryon-asymmetry
$Y_B=(n_B-n_{\bar{B}})/s=C Y_L$ ($C\approx -1/3$ for MSSM) by the
EW sphaleron effects. Using the Dirac and the Majorana
mass-matrices of Sec.~4, with the introduction of CP-violating
phases in them as discussed in Sec.~5, the lepton-asymmetry
produced per $N_1$ (or ($\tilde{N}_1+\bar{\tilde{N}}_1)$-pair)
decay is found to be \cite{PatiLepto}:
\begin{eqnarray}
\epsilon_{1} & \approx &
\frac{1}{8\pi}\left(\frac{\mathcal{M}_u^0}{v}\right)^2
|(\sigma+3\epsilon)-y|^2\sin\left(2\phi_{21}\right)\times (-3)\left(\frac{M_1}{M_2}\right)\nonumber\\
& \approx & -\left(2.0\times 10^{-6}\right)
\sin\left(2\phi_{21}\right) \times\left[\frac{(M_1/M_2)}{5\times
10^{-3}}\right] \label{eq:28}
\end{eqnarray}
\emph{Here $\phi_{21}$ denotes an effective phase depending upon
phases in the Dirac as well as Majorana mass-matrices (see
Ref.~\cite{PatiLepto}).} Note that the parameters $\sigma$,
$\epsilon$, $y$ and $(\mathcal{M}_u^0/v)$ are already determined
within our framework (to within 10 \%) from considerations of
fermion masses and neutrino oscillations (see Sec.~4 and 5).
Furthermore, from Eq.~(\ref{eq:MajM}) we see that $M_1\approx(1/3
- 3)\times 10^{10}$ GeV, and $M_2\sim 2\times10^{12}$ GeV, thus
$M_1/M_2\approx(5\times 10^{-3})(1/3 - 3)$. In short, leaving
aside the phase factor, the RHS of Eq.~(\ref{eq:28}) is pretty
well determined within our framework (to within about a factor of
5), as opposed to being uncertain by orders of magnitude either
way. \emph{This is the advantage of our obtaining the
lepton-asymmetry in conjunction with a predictive framework for
fermion masses and neutrino oscillations.} Now the phase angle
$\phi_{21}$ is uncertain because we do not have any constraint yet
on the phases in the Majorana sector $(M^\nu_R)$. At the same
time, since the phases in the Dirac sector are relatively large
(see Sec.~5 and Ref.~\cite{BPR}), barring unnatural cancellation
between the Dirac and Majorana phases, we would naturally expect
$\sin(2\phi_{21})$ to be sizable-i.e. of order $1/10$ to $1$
(say).

The lepton-asymmetry is given by $Y_L= \kappa (\epsilon_1/g^*)$,
where $\kappa$ denotes an efficiency factor representing wash-out
effects and $g^*$ denotes the light degrees of freedom
($g^*\approx 228$ for MSSM). For our model, using recent
discussions on $\kappa$ from Ref.~\cite{69BDB}, we obtain:
$\kappa\approx(1/18 - 1/60)$, for the thermal case, depending upon
the $''31''$ entries in the neutrino-Dirac and Majorana
mass-matrices (see Ref.~\cite{PatiLepto}). Thus, for the thermal
case, we obtain:
\begin{equation}
(Y_B)_{thermal}/\sin(2\phi_{21})\approx (10 - 30)\times 10^{-11}
\label{eq:29}
\end{equation}
where, for concreteness, we have chosen $M_1\approx 4\times 10^9$
GeV and $M_2\approx 1\times 10^{12}$ GeV, in accord with
Eq.~(\ref{eq:MajM}). In this case, the reheat temperature would
have to be about few $\times 10^9$ GeV so that $N_1$'s can be
produced thermally. We see that the derived values of $Y_B$ can in
fact account for the recently observed value $(Y_B)_{WMAP}\approx
(8.7 \pm 0.4)\times 10^{-11}$ \cite{70WMAP}, for a natural value
of the phase angle $\sin(2\phi_{21})\approx (1/3- 1)$. As
discussed below, the case of non-thermal leptogenesis can allow
even lower values of the phase angle. It also typically yields a
significantly lower reheat temperature ($\sim10^7 - 10^8$ GeV)
which may be in better accord with the gravitino-constraint.

For the non-thermal case, to be specific one may assume an
effective superpotential \cite{71Shafi}: $W^{infl}_{eff}=\lambda S
(\bar{\Phi}\Phi-M^2) +$ (non-ren. terms) so as to implement hybrid
inflation; here $S$ is a singlet field and $\Phi$ and $\bar{\Phi}$
are Higgs fields transforming as $(1, 2, 4)$ and $(1, 2, \bar{4})$
of $G(224)$ which break B-L at the GUT scale and give Majorana
masses to the RH neutrinos. Following the discussion in
\cite{71Shafi}, \cite{PatiLepto}, one obtains: $m_{infl}=\sqrt 2
\lambda M$, where $M=<(1, 2, 4)_H>\approx 2\times 10^{16}$ GeV;
$T_{RH}\approx(1/7)(\Gamma_{infl}M_{Pl})^{1/2}\approx(1/7)(M_1/M)
(m_{infl}M_{Pl}/8\pi)^{1/2}$ and
$Y_B\approx-(1/2)(T_{RH}/m_{infl}) \varepsilon_1$. Taking the
coupling $\lambda$ in a plausible range $(10^{-5} - 10^{-6})$
(which lead to the desired reheat temperature, see below) and the
asymmetry-parameter $\varepsilon_1$ for the
$G(224)/SO(10)$-framework as given in Eq.~(\ref{eq:28}), the
baryon-asymmetry $Y_B$ can then be derived. The values for $Y_B$
thus obtained are listed in Table~\ref{tab:1}.
\begin{table}
\begin{tabular}{|l|l|l|} \hline
$\lambda$ &  $10^{-5}$ &  $10^{-6}$ \\ \hline \hline $m_{infl}$
GeV &  $3\times 10^{11}$
    & $3\times 10^{10}$ \\ \hline
$T_{RH}$ GeV &  $(5.3-1.8)\times 10^7$ &
    $(17-5.6)\times 10^6$ \\ \hline
$\frac{Y_B\times 10^{11}}{\sin(2\phi_{21})}$ & $(100-10)$
    & $(300-33)$ \\ \hline
\end{tabular}
\caption{Baryon Asymmetry For Non-Thermal Leptogenesis}
\label{tab:1}
\end{table}

The variation in the entries correspond to taking $M_1=(2\times
10^{10} \mbox{ GeV})(1-1/3)$ with  $M_2=(2\times 10^{12})$ GeV in
accord with Eq.~(\ref{eq:MajM}). We see that for this case of
non-thermal leptogenesis, one quite plausibly obtains
\begin{eqnarray}
(Y_B)_{Non-Thermal}\approx(8 - 9)\times10^{-11} \end{eqnarray}

 \noindent in full accord with the WMAP data, for natural values of the phase
angle $\sin(2\phi_{21})\approx(1/3 - 1/10)$, and with $T_{RH}$
being as low as $10^7$ GeV $(2-1/2)$. Such low values of the
reheat temperature are fully consistent with the
gravitino-constraint for $m_{3/2}\approx 400$ GeV $- 1$ TeV (say),
even if one allows for possible hadronic decays of the gravitinos
for example via $\gamma\tilde{\gamma}$-modes \cite{72KKM}.

In summary, I have presented two alternative scenarios (thermal as
well as non-thermal) for inflation and leptogenesis. We see that
the $G(224)/SO(10)$-framework provides {\it a simple and unified
description} of not only fermion masses, neutrino oscillations
(consistent with maximal atmospheric and large solar oscillation
angles) \emph{and} CP violation, but also of baryogenesis via
leptogenesis, in either scenario. Each of the following features -
(a)~the existence of the RH neutrinos, (b)~B-L local symmetry,
(c)~$SU(4)$-color, (d)~the SUSY unification scale, (e)~the seesaw
mechanism, and (f)~the pattern of $G(224)/SO(10)$ mass-matrices
allowed in the minimal Higgs system (see Sec.~4)-have played
crucial roles in realizing this \emph{unified and successful
description}.

\section{Proton Decay}

Perhaps the most dramatic prediction of grand unification is
proton decay. I have discussed proton decay in the context of the
SUSY $SO(10)/G(224)$-framework presented here in some detail in
recent reviews \cite{JCPErice,JCPKeK} which are updates of the
results obtained in \cite{BPW}. Here, I will present only the
salient features and the updated results. In SUSY unification
there are in general three distinct mechanisms for proton decay.
\begin{enumerate}
\item \textbf{The familiar d=6 operators} mediated by $X$ and $Y$
gauge bosons of $SU(5)$ and $SO(10)$ As is well known, these lead
to $e^+\pi^0$ as the dominant mode with a lifetime $\approx
10^{35.3 \pm 1}$ yrs. \item \textbf{The ``standard'' $d=5$
operators} \cite{69SYW} which arise through the exchange of the
color-triplet Higgsinos which are in the $5_H +\bar{5}_H$ of
$SU(5)$ or $10_H$ of $SO(10)$. These operators require (for
consistency with proton lifetime limits) that the color-triplets
be made superheavy while the EW-doublets are kept light by a
suitable doublet-triplet splitting mechanism (for $SO(10)$, see
Ref.~\cite{DimWil}. They lead to dominant $\bar{\nu}K^+$ and
comparable $\bar{\nu}\pi^+$ modes with lifetimes varying from
about $10^{29}$ to $10^{34}$ years, depending upon a few factors,
which include the nature of the SUSY-spectrum and the matrix
elements (see below). In the present context, see
\cite{BPW,JCPKeK,JCPEriceReview}. Some of the original references
on contributions of standard $d=5$ operators to proton decay may
be found in
\cite{DimopRabyWilczek,Ellis,NathChemArno,Hisano,BabuBarr,LucasRaby,Murayama}
\item \textbf{The so called ``new'' $d=5$ operators}
\cite{BPW2,BPW} which can generically arise through the exchange
of color-triplet Higgsinos in the  Higgs multiplets like $(16_H
+\bar{16}_H)$ of $SO(10)$. Such exchanges are possible by
utilizing the joint effects of (a)~the couplings given in
Eq.~(\ref{eq:LMaj}) which assign superheavy Majorana masses to the
RH neutrinos through the VEV of $\bar{16}_H$, and (b)~the coupling
of the form $g_{ij} 16_i 16_j 16_H 16_H/M$ (see
Eq.~(\ref{eq:Yuk})) which are needed, at least for the minimal
Higgs-system, to generate CKM-mixings. These operators also lead
to $\bar{\nu}K^+$ and $\bar{\nu}\pi^+$ as the dominant modes, and
they can quite plausibly lead to lifetimes in the range of
$10^{32}-10^{34}$ yrs [see below]. These operators, though most
natural in a theory with Majorana masses for the RH neutrinos,
have been invariably omitted in the literature.
\end{enumerate}

One distinguishing feature of the new $d=5$ operator is that they
directly link proton decay to neutrino masses via the Majorana
masses of the RH neutrinos. \emph{The other, and perhaps most
important, is that these new $d=5$ operators can induce proton
decay even when the $d=6$ and standard $d=5$ operators mentioned
above are absent.} This is what could happen if the string theory
\cite{stringG(224)} or a higher dimensional GUT-theory
\cite{5DG(224)} leads to an effective $G(224)$-symmetry in $4D$,
which would be devoid of both $X$ and $Y$ gauge bosons and the
dangerous color-triplets in the $10_H$ of $SO(10)$. \emph{By the
same token, for an effective $G(224)$-theory, these new $d=5$
operators can become the sole and viable source of proton decay
leading to lifetimes in an interesting range (see below).}

Our study of proton decay carried out in Ref.~\cite{BPW} and
updated in \cite{JCPEriceReview} and \cite{JCPKeK} has a few
distinctive features: (i)~It is based on a \emph{realistic
framework for fermion masses and neutrino oscillations}, as
discussed in Sec.~4;(ii)~It includes the \emph{new $d=5$
operators} in addition to the standard $d=5$ and $d=6$ operators;
(iii)~It restricts \emph{GUT-scale threshold-corrections} to
$\alpha_3(m_Z)$ so as to be in accord with the demand of
``natural'' coupling unification and thereby restricts $M_{eff}$
that controls the strength of the standard $d=5$ operators; and
(iv)~It allows for the ESSM extension \cite{BabuJi} of MSSM
motivated on several grounds (see e.g. \cite{BabuJi} and
\cite{JCPEriceReview}), which introduces two vectorlike families
in $16+\bar{16}$ of $SO(10)$ with masses of order $1$ TeV, in
addition to the three chiral families.

Guided by recent calculation based on quenched lattice QCD in the
continuum limit \cite{81Tsutsui} and renormalization factors $A_L$
and $A_s$ for d = 5 as in \cite{Turznyski}, we take (see Ref.
\cite{JCPKeK} for details): $|\beta_H| \approx
|\alpha_H|\approx(0.009\mbox{ GeV}^3)(1/\sqrt{2}-\sqrt{2})$;
$m_{\tilde{q}}\approx m_{\tilde{l}}\approx 1.2$ TeV $(1/2-2)$;
$(m_{\tilde{W}}/m_{\tilde{q}})=1/6(1/2-2)$; $M_{H_C}(\mbox{min}
SU(5))\leq 10^{16}$ GeV, $A_L\approx 0.32$, $A_S\approx 0.93$,
$\tan{\beta}\leq 3$; $M_X\approx M_Y\approx 10^{16}$ GeV
$(1\pm25\%)$, and $A_R(d= 6, e^+\pi^0)\approx 3.4$.

The theoretical predictions for proton decay for the cases of
minimal SUSY $SU(5)$, SUSY $SO(10)$ and $G(224)$-models developed
in Secs.~3 and 4, are summarized in Table~3. They are obtained by
following the procedure as in \cite{BPW,JCPKeK,JCPEriceReview} and
using the parameters as mentioned above.\footnote{The chiral
Lagrangian parameter ($D+F$) and the renormalization factor
$A_{R}$ entering into the amplitude for $p \rightarrow
e^{+}\pi^{0}$ decay are taken to be 1.25 and 3.4 respectively.}
\begin{table*}
\centering
\begin{eqnarray}
\left. \frac{\mbox{SUSY $SU(5)$}}{\mbox{MSSM (std. $d=5$)}}
\right\}
\begin{array}{l}\Gamma^{-1}(p \rightarrow \bar{\nu}K^{+}) \end{array} & \leq &
\begin{array}{lr}
\begin{array}{l} 1.2 \times 10^{31} \mbox{ yrs} \end{array} &
\left( \begin{array}{c} \mbox{Excluded by} \\ \mbox{SuperK}
\end{array} \right)
\end{array} \label{new31}\\ \nonumber\\
\left. \frac{\mbox{SUSY $SO(10)$}}{\mbox{MSSM (std. $d=5$)}}
\right\}
\begin{array}{l} \Gamma^{-1}(p \rightarrow \bar{\nu}K^{+}) \end{array} & \leq &
\begin{array}{lr}
\begin{array}{l} 1 \times 10^{33} \mbox{ yrs} \end{array} &
\left( \begin{array}{c} \mbox{Tightly constrained} \\
\mbox{by SuperK} \end{array} \right)
\end{array} \label{new32} \\ \nonumber\\
\left. \frac{\mbox{SUSY $SO(10)$}}{\mbox{ESSM (std. $d=5$)}}
\right\}
\begin{array}{l}
\Gamma^{-1}(p \rightarrow \bar{\nu}K^{+})_{\mbox{Med.}} \\
\Gamma^{-1}(p \rightarrow \bar{\nu} K^{+})
\end{array}
&
\begin{array}{c} \approx \\ \lsim \end{array}
&
\begin{array}{lr}
\begin{array}{l}
(\mbox{1--10}) \times 10^{33} \mbox{ yrs} \\ 10^{35} \mbox{ yrs}
\end{array}
& \left( \begin{array}{c} \mbox{Fully SuperK} \\ \mbox{Compatible}
\end{array} \right)
\end{array} \label{new33} \\ \nonumber\\
\left. \frac{\mbox{SUSY $G(224)/SO(10)$}}{\mbox{MSSM or ESSM (new
$d=5$)}} \right\}
\begin{array}{l}
\Gamma^{-1}(p \rightarrow \bar{\nu}K^{+}) \\
B(p \rightarrow \mu^{+} K^{0})
\end{array}
&
\begin{array}{c} \lsim \\ \approx \end{array}
&
\begin{array}{lr}
\begin{array}{l}
2 \times 10^{34} \mbox{ yrs} \\ (1-50)\%
\end{array}
& \left( \begin{array}{c} \mbox{Fully Compatible} \\ \mbox{with
SuperK}
\end{array} \right)
\end{array} \label{new34} \\ \nonumber\\
\left. \frac{\mbox{SUSY $SU(5)$ or $SO(10)$}}{\mbox{MSSM ($d=6$)}}
\right\}
\begin{array}{l} \Gamma^{-1}(p \rightarrow e^{+} \pi^{0}) \end{array} & \approx
& \begin{array}{lr} \begin{array}{l} 10^{35 \pm 1} \mbox{ yrs}
\end{array} & \left( \begin{array}{c} \mbox{Fully Compatible} \\
\mbox{with SuperK}
\end{array} \right)
\end{array} \label{new35}
\end{eqnarray}
\caption{A Summary of Results on Proton Decay}
\end{table*}

It should be stressed that the upper limits on proton lifetimes
given in Table~3 are quite conservative in that they are obtained
(especially for the top two cases) by stretching the uncertainties
in the matrix element and the SUSY spectra to their extremes so as
to prolong proton lifetimes.  In reality, the lifetimes should be
shorter than the upper limits quoted above.

Now the experimental limits set by SuperK studies are as follows
\cite{SKlimit}:
\begin{eqnarray}
\label{proton}
 \Gamma^{-1}(p\rightarrow
e^{+}\pi^{0})_\mathrm{expt} & \geq &
    6 \times 10^{33}\mbox{ yrs} \nonumber \\
\Gamma^{-1}(p\rightarrow \bar{\nu}K^{+})_{\mathrm{expt}} & \geq &
    1.9 \times 10^{33}\mbox{ yrs}
\end{eqnarray}
The following comments are in order.
\begin{enumerate}
\item By comparing the upper limit given in Eq.~(\ref{new31}) with
the experimental lower limit, we see that the \emph{minimal} SUSY
$SU(5)$ with the conventional MSSM spectrum is clearly excluded by
a large margin by proton decay searches.  This is in full
agreement with the conclusion reached by other authors (see e.g.
Ref.~\cite{Murayama}).\footnote{See, however,
Refs.~\cite{BajcPerezSenja} and \cite{EmmanuelWies}, where
attempts are made to save minimal SUSY SU(5) by a set of
scenarios. These include a judicious choice of sfermion mixings,
higher dimensional operators and squarks of first two families
having masses of order 10 TeV.} \item By comparing
Eq.~(\ref{new32}) with the empirical lower limit, we see that the
case of MSSM embedded in $SO(10)$ is already tightly constrained
to the point of being disfavored by the limit on proton lifetime.
The constraint is of course augmented by our requirement of
\emph{natural coupling unification}, which prohibits accidental
large cancelation between different threshold corrections (see
\cite{BPW}). \item In contrast to the case of MSSM, that of ESSM
\cite{BabuJi} embedded in $SO(10)$, which has been motivated on
several grounds\footnote{The case of ESSM, which introduces two
vector like families, i.e. $16 + \overline{16}$ of SO(10), with a
mass of order 1 TeV, has been motivated by a number of
considerations independently of proton decay \cite{BabuJi}. These
include: (a) dilaton stabilization through a semi-perturbative
unification, (b) coupling unification with a better prediction for
$\alpha_3(m_Z)$ compared to that for MSSM, (c) a simple
understanding of the inter-family mass hierarchy, and (d) a
possible explanation of a 2.7 $\sigma$ anomaly in $(g-2)_\mu$. The
vector like families with mass of order 1 TeV can of course be
searched for at the LHC.}, is fully compatible with the SuperK
limit (see Eq.~(\ref{new33})). In this case,
\(\Gamma_{\mathrm{Med}}^{-1}(p \rightarrow \bar{\nu}K^{+}) \approx
10^{33} - 10^{34} \mbox{ yrs}\), given in Eq.~(\ref{new33}),
corresponds to the parameters involving the SUSY spectrum and the
matrix element $\beta_{H}$ being in the \emph{median range}, close
to their central values. \item We see from Eq.~(\ref{new34}) that
the contribution of the new operators \cite{BPW2} related to the
Majorana masses of the RH neutrinos (which is the same for MSSM
and ESSM and is independent of $\tan \beta$) is fully compatible
with the SuperK limit.  These operators can quite naturally lead
to proton lifetimes in the range of $10^{33}-10^{34}$ yrs with an
upper limit of about \(2 \times 10^{34}\) yrs.
\end{enumerate}

In summary for this section, within the $SO(10)/G(224)$ framework
and with the inclusion of the standard as well as the new $d=5$
operators, one obtains (see Eqs.~(\ref{new32})--(\ref{proton})) a
conservative upper limit on proton lifetime given by:
\begin{equation}
\begin{array}{lr}
\tau_{\mathrm{proton}} \lsim (1/3 - 2) \times 10^{34}
\mbox{ yrs} & \left( \begin{array}{c} \mbox{SUSY} \\
SO(10)/G(224) \end{array} \right)
\end{array}
\label{new36}
\end{equation}
with $\bar{\nu}K^{+}$ and $\bar{\nu}\pi^{+}$ being the dominant
modes and quite possibly $\mu^{+}K^{0}$ being prominent.

The $e^+\pi^0$-mode induced by gauge boson-exchanges should have
an inverse decay rate in the range of $10^{34}-10^{36}$ years (see
Eq. (35)).  The implication of these predictions for a
next-generation detector is noted in the next section.

\section{Concluding Remarks}

The neutrinos seem to be as elusive as revealing. Simply by virtue
of their tiny masses, they provide crucial information on the
unification-scale, and even more important on the nature of the
unification-symmetry. In particular, as argued in Secs.~4 and 6,
(a)~the magnitude of the superK-value of $\sqrt{\delta m^2_{23}}
(\approx 1/20 \mbox{ eV})$, (b)~the $b/\tau$ mass-ratio, and
(c)~the need for baryogenesis via leptogenesis, together, provide
clear support for: (i)~the existence of the $SU(4)$-color symmetry
in 4D above the GUT-scale which provides not only the RH neutrinos
but also B-L as a local symmetry and a value for
$m(\nu^{\tau}_{\mbox{Dirac}})$; (ii)~the familiar SUSY
unification-scale which provides the scale of $M_R$; and (iii)~the
seesaw mechanism. \emph{In turn this chain of arguments selects
out the effective symmetry in $4D$ being either a string-derived
$G(224)$ or $SO(10)$-symmetry, as opposed to the other
alternatives like $SU(5)$ or flipped $SU(5)'\times U(1)$}.

It is furthermore remarkable that the tiny neutrino-masses also
seem to hold the key to the origin of baryon excess and thus to
our own origin!

In this talk, I have tried to highlight that the
$G(224)/SO(10)$-framework as described here is capable of
providing a \emph{unified description} of a set of phenomena
including: fermion masses, neutrino oscillations, CP and flavor
violations as well as of baryogenesis via leptogenesis. This seems
non-trivial.

The neutrinos have clearly played a central role in arriving at
this unified description, first (a) by providing a clue to the
nature of the unification-symmetry (as noted above), second (b) by
confirming certain group-theoretic correlations between the quark
and lepton sectors as regards their masses and mixings (cf.
$m(\nu_\tau)_{\rm Dirac}$ versus $m_{top}$ and
$\theta^{osc}_{\nu_\mu \nu_\tau}$ versus $V_{cb}$), and (c) by
yielding naturally the desired magnitude for the baryon excess.
Hence the title of the paper.

The framework is also highly predictive and can be further tested
by studies of CP and flavor violations in processes such as
(a)~$B_d\rightarrow\phi K_S$-decay, (b)~$(B_S,\bar{B}_S)$-decays,
(c)~edm of neutron, and (d)~leptonic flavor violations as in
$\mu\rightarrow e\gamma$ and $\tau\rightarrow\mu\gamma$-decays,
and in $\mu N\to e N$.

To conclude, the evidence in favor of supersymmetric grand
unification, based on a string-derived $G(224)$-symmetry in 4D (as
described in Sec.~3) or $SO(10)$-symmetry, appears to be strong.
It includes:
\begin{itemize}
\item Quantum numbers of all members in a family, \item
Quantization of electric charge, \item Gauge coupling unification,
\item $m^0_b\approx m^0_\tau$ \item $\sqrt{\delta
m^2(\nu_2-\nu_3)}\approx 1/20$ eV, \item A maximal
$\Theta^{\nu}_{23}\approx \pi/4$ with a minimal $V_{cb} \approx
0.04$, and \item Baryon Excess $Y_B\approx 10^{-10}$.
\end{itemize}
All of these features and more including (even) CP and flavor
violations hang together neatly {\it within a single unified
framework} based on a presumed string-derived four-dimensional
$G(224)$ or $SO(10)$-symmetry, with supersymmetry. It is hard to
believe that this neat fitting of all these pieces emerging as
predictions of one and the same framework can be a mere
coincidence. It thus seems pressing that dedicated searches be
made for the two missing pieces of this picture-that is
supersymmetry and proton decay. The search for supersymmetry at
the LHC and a possible future NLC is eagerly awaited. That for
proton decay will need a next-generation megaton-size underground
detector.

\textbf{Acknowledgments}: I would like to thank Milla
Baldo-Ceolini for her kind hospitality. I have benefited from many
collaborative discussions with Kaladi S. Babu, Parul Rastogi and
Frank Wilczek on topics covered in this lecture.  Conversations
with Milla Baldo-Ceolini, Manfred Lindner, Antonio Masiero and
Qaisar Shafi during the Venice conference leading to bets on some
interesting issues in particle physics, which will be settled soon
by experiments, were enlightening. The research presented here is
supported in part by DOE grant No.~DE-FG02-96ER-41015.


\begin{thebibliography}{29}
\bibitem{FN4} The zeros in ``11", ``13" and ``31" elements signify that they
are relatively small quantities (specified below). While the ``22"
elements were set to zero in Ref. \cite{BPW}, because they are
meant to be $<$``23" ``32"/``33"$\sim 10^{-2}$ (see below), and
thus unimportant for purposes of Ref. \cite{BPW}, they are
retained here, because such small $\zeta_{22}^u$ and
$\zeta_{22}^d$ [$\sim (1/3)\times 10^{-2}$ (say)] can still be
important for CP violation and leptogenesis.
\bibitem{FN26}
For G(224), one can choose the corresponding sub-multiplets --
that is (1, 1,
  15)$_H$, (1, 2, $\bar{4}$)$_H$, (1, 2, 4)$_H$, (2, 2, 1)$_H$ -- together with
  a singlet $S$, and write a superpotential analogous to Eq. (\ref{eq:Yuk}).
\end{thebibliography}
\end{document}